\documentclass{jfm}
\usepackage{graphicx}
\usepackage{epstopdf, epsfig}
\usepackage{amsmath,amssymb}
\usepackage{natbib}
\usepackage[mathlines]{lineno}
\usepackage[dvipsnames]{xcolor}
\usepackage{tikz}
\usepackage{mathtools}
\usepackage{color}
\usepackage{changes}

\usepackage{epstopdf} %converting to PDF
\usepackage{appendix}

\makeatletter
\newcommand{\specialnumber}[1]{%
	\def\tagform@##1{\maketag@@@{(\ignorespaces##1\unskip\@@italiccorr#1)}}}
\newcommand{\specialeqref}[2]{\begingroup
	\def\tagform@##1{\maketag@@@{(\ignorespaces##1\unskip\@@italiccorr#2)}}%
	\eqref{#1}\endgroup}

\newcommand{\ed}[1]{{\color{black}#1}}

\newcommand{\eps}{\epsilon}
\newcommand{\mo}{\mathcal{O}}
\newcommand{\rmi}{\mathrm{i}}
\newcommand{\rme}{\mathrm{e}}
\newcommand{\rmd}{\mathrm{d}}

\newcommand{\cc}{\text{c.c.}}
\newcommand{\be}{\mathbf{e}}

\newcommand{\brW}{\bar{W}}
\newcommand{\bx}{\mathbf{x}}

\newcommand{\bk}{\mathbf{k}}

\newcommand{\bkp}{\pmb{\kappa}}

\newcommand{\hzt}{\hat{\zeta}}

\newcommand{\half}{\dfrac{1}{2}}
\newcommand{\quard}{\dfrac{1}{4}}

\newcommand{\brw}{\bar{w}}
\newcommand{\brwc}{\bar{W}}

\newcommand{\mR}{\mathcal{R}}
\newcommand{\mn}{\mathcal{N}}
\newcommand{\mt}{\mathcal{T}}

\newcommand{\mw}{\mathcal{W}}
\newcommand{\mj}{^{(mj)}}

\newcommand{\bmw}{\bar{\mw}}
\newcommand{\bmt}{\bar{\mt}}

\newcommand{\bme}{\pmb{\mathcal{E}}}

\newcommand{\upj}{^{[j]}}
\newcommand{\tf}{\text{f}}

\newcommand{\kx}{\bk\cdot\bx}
\newcommand{\kint}{\int\limits_{-\infty}^{\infty}}
\newcommand{\efackx}{\rme^{\rmi \kx}} 

\newcommand{\efac}{\rme^{\rmi (\alpha \bk_0\cdot\bx-\beta\omega_0t)} }
\newcommand{\efacj}{\rme^{\rmi j (\alpha \bk_0\cdot\bx-\beta\omega_0t)} }

\newcommand{\efact}{\rme^{2\rmi  (\alpha \bk_0\cdot\bx-\beta\omega_0t)} }
\newcommand{\efactrd}{\rme^{3\rmi (\alpha \bk_0\cdot\bx-\beta\omega_0t)} }

\newcommand{\efacff}{\rme^{\rmi ( \bk_0\cdot\bx-\omega_0t +\theta_0)} }
\newcommand{\efactf}{\rme^{2\rmi ( \bk_0\cdot\bx-\omega_0t+\theta_0)} }

\newcommand{\etmj}{\rme^{-\rmi j\beta\omega_0t} }

\newcommand{\fst}{^{(1)}}
\newcommand{\snd}{^{(2)}}
\newcommand{\sz}{^{(20)}}
\newcommand{\ssd}{^{(22)}}
\newcommand{\trdf}{^{(31)}}
\newcommand{\trd}{^{(3)}}
\newcommand{\trdt}{^{(33)}}
\shorttitle{CEEEs for nonlinear surface waves}
\shortauthor{Li}

\title[CEEEs for nonlinear surface gravity waves]{On coupled envelope evolution equations in the Hamiltonian theory of nonlinear surface gravity waves}

\author{ 
	Yan Li\aff{1}
	\corresp{\email{yan.li@uib.no}}, 
}
\affiliation{
	\aff{1} Department of Mathematics, University of Bergen, N-5020 Bergen, Norway
}

\begin{document}
% \linenumbers
	\maketitle
	\begin{abstract}
		This paper presents a novel theoretical framework in the Hamiltonian theory of nonlinear surface gravity waves. The envelope of surface elevation and the velocity potential on the free water surface are introduced in the framework, which are shown to be a new pair of canonical variables. Using the two envelopes as the main unknowns, coupled envelope evolution equations (CEEEs) are derived  based on a perturbation expansion. Similar to the High Order Spectral method, the CEEEs can be derived up to arbitrary order in wave steepness. In contrast, they have a temporal scale as slow as the rate of change of a wave spectrum and allow for the wave fields prescribed on a computational (spatial) domain with a much larger size and with spacing longer than the characteristic wavelength at no expense of accuracy and numerical efficiency. The energy balance equation is derived based on the CEEEs. The nonlinear terms in the CEEEs are in a form of the separation of wave harmonics, due to which an individual term is shown to have clear physical meanings in terms of whether or not it is able to force free waves which obey the dispersion relation. Both the nonlinear terms that can only lead to the forcing of bound waves and these which are capable of forcing free waves are demonstrated, with the latter through the analysis of the quartet and quintet resonant interactions of linear waves. The relations between the CEEEs and two other existing theoretical frameworks are established, including the theory for a train of Stokes waves up to second order in wave steepness [Fenton, \textit{J. waterway, Port, Coast. \& Ocean Eng.}, \textbf{111}, 2, 1985] and a semi-analytical framework for three-dimensional weakly nonlinear surface waves with arbitrary bandwidth and large directional spreading by
		Li \& Li [\textit{Phys. Fluids}, \textbf{33}, 7, 2021]. 

	\end{abstract}
	
	%\begin{keywords}
	%waves/free-surface flow, surface water waves, geophysical and geological flows,  coastal engineering.
	% Authors should not enter keywords on the manuscript, as these must be chosen by the author during the online submission process and will then be added during the typesetting process (see https://www.cambridge.org/core/services/aop-file-manager/file/577245f575fc11ff73cc3197/jfm-keywords.pdf
	%\end{keywords}
	
%-------------%---%----%----%----%--------------%
%                    section                    %
%-------------%---%----%----%----%--------------%
%
\section{Introduction}
For the description of non-breaking surface water waves in the open ocean and coastal waters, there are many available approaches in the framework of potential flow. The high-order spectral (HOS) method \citep{dommermuth87,west87},  Boussinesq-type formulations \citep{WEI95,agnon99},  volumetric
methods \citep{engsig09,bihs20}, and the fast computational method developed by \cite{clamond01} are a few examples of these which can account for waves up to arbitrary order in wave steepness. In terms of the numerical efficiency for a given level of accuracy, it is without doubt that the HOS method is the preferred choice compared with the aforementioned alternatives \citep{klahn20}. Two aspects have especially contributed to its high efficiency: (i) it permits an explicit method for the time integration and vertical velocity on the free water surface and (ii) it takes advantage of spectral methods for numerical computations \citep{dommermuth87,ducrozet16}. 

In cases for the evolution of weakly nonlinear waves,  a nonlinear Schr\"{o}dinger (NLS) equation-based model \citep{benney67,zakharov68,chu71,davey74,dysthe179,trulsen96,trulsen00,li21nls} as well as the Hasselmann/Zakharov integral equation (Hasselmann 1962; Zakharov 1968;45
Janssen 1983; Stiassnie 1984; Krasitskii 1994) have
 been widely known as a powerful analytical tool. The NLS equation-based model is especially superior to the HOS method in the sense of the computational efficiency, arising from that it describes the evolution of an envelope which varies slowly in time and depends on a long length scale compared with the rapidly varying wave phase with a short length scale. A NLS equation, e.g., these by \cite{trulsen00,gramstad11a,li21nls}, can efficiently resolve the phase of free waves in a large computational domain while it well captures the wave energy transfers due to quartet interactions of waves within a narrow bandwidth. However,  approximations in addition to a perturbation expansion are necessary throughout the derivations of a NLS equation, having limited its wide validity  (see, e.g., p.202 by \cite{janssen04}). 
This especially indicates that their capability of accounting for the physics due to moderately nonlinear and steeper waves is likely compromised. 

\ed{Similar to a NLS-based equation, the reduced Zakharov equation has been widely used for elucidating the nonlinear physical properties of water waves, see, e.g., \cite{crawford80,janssen07,stiassnie09,gramstad14} among others. It produces explicit expressions for  the interaction of a number of up to five waves and its numerical  and theoretical potential has been extensively explored in recent years \citep{annenkov01,annenkov06,annenkov09,dyachenko11,dyachenko17}. One distinctive feature of the reduced Zakharov equation and its compact form is that they describe the evolution of a complex function which is defined using the Hamiltonian structure of physical variables for eliminating non-resonant interaction terms. This means other wave fields such as the surface elevation and velocity are evaluated at an additional cost. Nevertheless, existing wave-averaged equations of ocean mean flows commonly rely on the input of the wave-induced (Eulerian or Lagrangian) velocity and surface elevation (see, e.g., \cite{sullivan10,suzuki16} and references therein). In such a context, the use of the new complex function may not necessarily introduce the advantage of the possible elimination of the nonlinear non-resonant terms for surface waves in a non-conservative system as they are expected to play a role. Thus, the Zakharov equation and its compact and numerical versions  are not seemingly ideal as a wave-phase resolved coupled model in large-scale physical processes in the open ocean, either. }

The superior features of a NLS equation-based model to the HOS method are especially important in the study of the roles of surface waves in the dynamics of the upper ocean, e.g., vertical mixing and the circulation of submesoscale currents, attributing to two aspects. Firstly, various important physical processes, e.g., the exchange of the momentum and energy flux between surface waves and a submesoscale flow, occur in a temporal and length scale at a magnitude significantly larger than that of surface waves, as has been especially found in wave averaged equations for the dynamics of the upper ocean flows \citep{mcwilliams04,sullivan10,suzuki16}. This suggests the need for capturing the long-term energy evolution of waves described on an extremely large domain. Secondly, recent studies find that wave phases play an important role in distorting turbulence with a scale smaller than the surface waves \citep{teixeira02,thorpe04,dasaro14} and the generation of turbulence by non-breaking surface waves \citep{babanin06,benilov12}, addressing the additional need for resolving wave phases for important physical mechanisms with a small scale. 

Indeed, a NLS equation-based model has been included in the framework of Regional Ocean Modeling System (ROMS) based on \cite{mcwilliams04}, despite that a few important nonlinear wave physics such as the Benjamin-Feir instability \citep{benjamin67,longuet78,janssen09} and quartet resonant interaction of waves \citep{phillips60} have not been considered yet. Moreover, due to the multiple scales involving a few orders of magnitude, the understanding of the coupled effects between small-scale turbulence, middle-scale surface waves, and large-scale submesoscale currents have been extremely limited. To make a difference, the author believes that it relies on an accurate and efficient model of surface waves, which should have the potential of bridging the connections between the smaller and larger scale physical processes with the scale in the middle being characterized by  surface waves. To this end, neither the HOS method nor a NLS equation-based model is seemingly ideal, the former of which due to the relatively low numerical efficiency for wave parameters required on an extremely large domain and the latter of which due to  the restricted accuracy and validity. % \ed{The numerical implementation of the HOS method requires a refined rectangular grid and a small time step for numerical convergence, implying a relatively low numerical efficiency for the long-term evolution of waves on an extremely large domain,  as discussed in \cite{annenkov01}.} 

Following the above discussion, an obvious question is whether it is possible to derive a framework which combines the advantages of both the HOS method and a NLS equation-based model, with  the potential of being applied in more general works which directly bridge the coupled physical processes of ocean surface waves with small-scale turbulence and submesoscale current. It means that such a framework should be as accurate as the HOS method while permitting the main numerical features of a NLS equation-based model. Specifically, it is desired to include the computational efficiency which allows for a large and coarse computation domain on which both the amplitude and wave phase can be well resolved.  Addressing this question defines the primary objective of this paper. It aims to present a new framework originally inspired by a NLS equation-based model in the manner that envelopes are introduced as the starting point.  The coupled envelope evolution equations  which can be derived accurate to arbitrary order in wave steepness are presented in the Hamiltonian theory. It should be noted that the idea of the envelope equations, which takes the advantages of both Fourier transforms and a newly defined linear operator, is different from that of the localized Zakharov equation (LZE). The idea of LZE deals with wave field dynamics in a manner of multiple interacting wave packets, posing challenges in its  mathematical implementations \citep{rasmussen99,gramstad11b}. \ed{With a need for extension, the newly derived framework is expected to especially have wide applicability in terms of coupling the surface waves-driven processes with regional oceanic dynamics in the upper ocean. }

This paper is laid out as follows. The statement of the problem is presented in \S\ref{sec:problem_Def}, followed by a review of the HOS method (\S\ref{sec:reviewHOS}) and a traditional perturbation method (\S\ref{sub:tra_pertb}) in \S\ref{sec:review}. A new framework proposed by this paper is presented in \S\ref{sec:ceees}. Especially in \S\ref{sub:newVar}, the (slowly spatial-temporal varying) envelope of both surface elevation and velocity potential on the free water surface -- which are firstly introduced by this paper, are shown to be a new pair of canonical variables. The detailed derivations for the Coupled Envelope Evolution Equations (CEEEs) are presented in \S\ref{sub:3methods}--\S\ref{sub:ceees}. A few important features of the CEEEs are explored in \S\ref{sec:ceeesFeatures} and \S\ref{sec:comprisons}. Three aspects of the newly derived CEEEs have been discussed in \S\ref{sec:ceeesFeatures}. How to apply an exponential integrator with the CEEEs is presented in \S\ref{sub:expInt}, together with the numerical implementation of the CEEEs which leads to the accurate description of linear waves. It is shown in \S\ref{sub:wae} that the CEEEs can  lead to the energy balance equation. In the limiting cases where wave nonlinearity can be neglected, the  energy balance equation is shown to naturally conserve the energy, implying no exchange of energy between linear waves as it should be. The nonlinear forcing terms in the CEEEs have clear physical meanings which are discussed in \S\ref{sec:nonLwaves}, including these which can only contribute to the forcing of bound waves and these which are capable of forcing free waves arising from the quartet and quintet resonant interactions of waves. The CEEEs are compared with a traditional perturbation method in \S\ref{sub:comp_pertb}, where the relations between the two methods are analytically shown for the evolution of a train of both Stokes waves \citep{fenton85} and three-dimensional waves with arbitrary bandwidth and with large directional spreading \citep{lili21}.  The comparisons between the CEEEs and the HOS method are discussed in \S\ref{sub:hoscomp} in numerical computations for  a limiting case and the numerical performances illustrated through a few numerical algorithms used for numerical implementations. The main conclusions drawn in this paper are presented in \S\ref{sec:conclud}. 
%

%-------------%---%----%----%----%--------------%
%                    section                    %
%-------------%---%----%----%----%--------------%
%
\section{Mathematical formulation}
\label{sec:problem_Def}
% {\color{red} How about adding the surface tension?}
%

\begin{table}
	\caption{Nomenclature	}
	\centering
	\begin{tabular}{ll} 
       \hline
      $(\bx,z)$ & (position vector in the horizontal space, vertical axis)
      \\
      $t$ & time
      \\
      $\rho$ & water density
      \\
      $g$ & gravitational acceleration 
      \\
      $h$ & water depth 
      \\
      $\bk_0$ ($k_0=|\bk_0|$) and $\omega_0$ & characteristic wave vector (wavenumber) and angular frequency \\
      $k_s$ & the wavenumber of the shortest wave that can be represented
      \\
      {$f_s$} & the frequency of the shortest wave that can be represented
      \\
      $\Delta t$ & the time interval 
      \\
      $\alpha, \beta$  & non-negative dimensionless parameters
      \\
      $M$ &  the truncated order of accuracy in wave steepness 
      \\
      $N_s$ &  the total number of discrete points used in a computational domain 
      \\
      \hline
      $\eps$ & dimensionless wave steepness used in the HOS method and CEEEs \\
      $\eps_0$ & non-dimensional wave steepness defined in \S\ref{sub:tra_pertb} \\
      $\varepsilon_t=f_s\Delta t$ & dimensionless time interval
      \\
      $\varepsilon_\tf=(f_s-\beta M f_0)/f_s$ & dimensionless frequency bandwidth
      \\
      $\varepsilon_k=(k_s-\alpha M k_0)/k_s$ & dimensionless wavenumber bandwidth
      \\
      %%%%%%%%%
      \hline
      $\zeta(\bx,z,t) $   & surface elevation
      \\
      $\psi(\bx,t)$ & velocity potential on the free water surface
      \\
      $W(\bx,t)$ & velocity  on the free water surface
      \\
	 $\Phi(\bx,z,t)$	& velocity potential
      \\
      $w(\bx,z,t)$ & vertical velocity
      \\
      %%%%%%%%%
      \hline
      $A(\bx,t)$ & The envelope of the surface elevation $\zeta$
      \\
      $B_s(\bx,t)\equiv B_0\equiv B_0^{(11)}$ & the envelope of the velocity potential ($\psi$) on the free water surface 
      \\
      $B(\bx,z,t)\equiv B^{(11)}$ & the envelope of the velocity potential $\Phi$
      \\
      $\brw(\bx,z,t)$ &  the envelope of the  vertical component of velocity $w$
      \\
      $\brwc(\bx,t)$ & the envelope of the  vertical velocity on the free water surface  $W$
      \\
      %%%%%%%%%
      \hline 
      subscript `0' & the evaluation at the still water surface $z=0$
      \\
      subscript `$M$'  &  the truncated order of accuracy in wave steepness
      \\
      subscripts `{$mj$}' & the $j$-th harmonic in the $m$-th order in wave steepness $\eps_0$
      \\
      superscripts `$(m)$' & the $m$-th order in wave steepness $\eps$ 
      \\
      superscripts `$(mj)$' & the $j$-th harmonic in the $m$-th order in wave steepness $\eps$
      \\
      \hline
      $\mw(\bx,t)$ and $\mt(\bx,t)$ & the nonlinear forcing term defined in the kinematic and dynamic    \\ & boundary conditions on the free water surface, respectively
      \\
      $\mn_A(\bx,t)$ and $\mn_B(\bx,t)$ & the nonlinear forcing terms newly introduced in the CEEEs
      \\
      \hline
      $\hat{(...)}$ & The Fourier transform of an arbitrary parameter $(...)$ with respect  \\
      & to the horizontal position vector into the Fourier $\bk$ space 
      \\
      the prime (')  & the fields only used in a traditional method presented in \S\ref{sub:tra_pertb}
\end{tabular}
\label{tab:non}
\end{table}

\subsection{Problem definition}
We consider ocean surface waves propagating on waters of a finite depth in the framework of potential-flow theory, thereby assuming incompressible inviscid flows and irrotational fluid motions, and negligible effects of surface tension.  A Cartesian coordinate system is chosen with the undisturbed water surface located at $z = 0$. A list of the main symbol notations used in this work is given in table \ref{tab:non}. 
The system  can  be described as a boundary value problem governed by the Laplace equation:
\begin{linenomath}
\begin{equation}
	\nabla^2_3 \Phi =0~~\textrm{for}~~ -h< z< \zeta(\bx,t),
	\label{eq:gvn_lap}     
\end{equation}
where $ \Phi(\bx,z,t) $ denotes the velocity potential, $ \zeta(\bx,t) $ is the free surface elevation,  $\bx$ is the position vector in the horizontal plane, $h$ is the water depth assumed to be constant, $t$ is the time, and $\nabla_3 = (\nabla,\p_z)$ with $\nabla=(\p_x,\p_y)$ denoting the gradient in the horizontal plane. Equation \eqref{eq:gvn_lap} should be solved subject to the nonlinear kinematic and dynamic boundary conditions (cf. \citet{davey74}) on the free water surface $z= \zeta(\bx,t)$, respectively
\begin{equation} \label{eq:gvn_sf}
	\p_t\zeta +\nabla \Phi\cdot \nabla\zeta - \p_z\Phi =  0
	~\textrm{and}~
	\p_t\Phi +g\zeta+\half(\nabla_3\Phi\big)^2=0,
	\specialnumber{a,b}
\end{equation}
\end{linenomath}
where $g$ denotes the gravitational acceleration;  A seabed boundary condition:
\begin{align} \label{eq:bc_sb}
	\p_z \Phi % \dl{+ \nabla \Phi\cdot\nabla h }
     ~= 0~~\textrm{for}~~z =-h,
\end{align}
where a constant uniform water depth $h$, is assumed. It should be noted that the extension of the new envelope equations presented in \S\ref{sec:ceees} to permit a slowly varying water depth with the horizontal position would be straightforward following \cite{dommermuth87} and the detailed derivations in this paper.

%---%---%----%----%----%
%              section             %
%---%---%----%----%----%
%
\subsection{Boundary conditions on the free water surface}
\label{sub:bcsf_Ham}
Following \cite{zakharov68} and \cite{krasitskii94}, we introduce the potential ($\psi$) and vertical velocity ($W$) defined on the unknown free water surface $z=\zeta(\bx,t)$ as follows,
\begin{linenomath}
\begin{equation}\label{eq:defs_psiW}
\psi(\bx,t) = \Phi(\bx, \zeta(\bx,t), t) ~ \text{and}~
W(\bx,t) = \p_z\Phi(\bx, z, t) ~\text{for}~z= \zeta(\bx,t).
\specialnumber{a,b}
\end{equation}
Inserting the definition of $\psi$ and $W$ into \specialeqref{eq:gvn_sf}{a,b} leads to the boundary conditions on the free water surface expressed as equations for unknowns $\psi(\bx,t)$, $W(\bx,t)$, and $\zeta(\bx,t)$, given by
\begin{equation}
\label{eq:bcnew_sf}
\p_t\zeta -W =-\nabla \psi\cdot \nabla\zeta +W(\nabla\zeta)^2~\text{and}~
\p_t\psi + g\zeta = - \half(\nabla\psi)^2+\half W ^2\big[1+(\nabla\zeta)^2\big].
\specialnumber{a,b}
\end{equation}
\end{linenomath}
The system described by the equations composed of \eqref{eq:gvn_lap}, \eqref{eq:bc_sb}, and \specialeqref{eq:bcnew_sf}{a,b} is known as the fully nonlinear (potential flow) boundary value problem in a Hamiltonian theory (see, e.g., \cite{west87}). An approximate solution to this problem can be obtained by using various methods as noted in the introduction, e.g., a High-Order Spectral (HOS) method \citep{dommermuth87,west87}, Hasselmmann/Zakharov integral equation \citep{hasselmann62,zakharov68,krasitskii94}, and the Coupled Envelope Evolution Equations (CEEEs) which are derived for the first time in \S\ref{sec:ceees} in this paper. 
%

%---%---%----%----%----%
%       section        %
%---%---%----%----%----%
%
\subsection{Velocity potential}
\label{sub:pertb_Ham}
We seek the solution for unknown potential ($\Phi(\bx,z,t)$)  of the Laplace equation  and the seabed boundary condition given by \eqref{eq:gvn_lap} and \eqref{eq:bc_sb}, respectively, in a form of power series in wave steepness denoted by $\eps$ which stands for a small nondimensional scaling parameter:
\begin{linenomath}
\begin{equation} \label{eq:phizeta_pertb}
   \Phi(\bx,z,t) = \sum_{m=1}^{M} \eps^m\Phi^{(m)}(\bx,z,t),
%   ~ \text{and}~
%   \zeta(\bx) = \sum_{m=1}^{M} \eps^m\zeta^{(m)}(\bx,t),
%   \specialnumber{a,b}
\end{equation}
where the terms are kept up to the $M-$th order in wave steepness and the superscript `($m$)' denotes $\mo(\eps^m)$, and the unknown (real) potential at the $m-$th order in wave steepness is given by (see, e.g., \cite{dommermuth87})
\begin{align} \label{eq:phij_ifft}
   \Phi^{(m)}(\bx,z,t)  = 
            \kint \hat{\Phi}_0^{(m)}(\bk,t)
            \dfrac{\cosh |\bk|(z+h)}{\cosh |\bk|h}\rme^{\rmi\kx}\rmd\bk,
\end{align}
\end{linenomath}
where $\hat{\Phi}^{(m)}_0(\bk,t)=\hat{\Phi}^{(m)}(\bk,0,t)$ denotes the $m-$th order velocity potential evaluated at $z=0$ in the Fourier $\bk$ plane; the subscript '0' is used to denote the evaluation at a still water surface $z=0$.

%---%---%----%----%----%
%       section        %
%---%---%----%----%----%
%
\subsection{Definition of two operators}
\label{sub:2operators}
\begin{figure}
	\centering
	\includegraphics[width=.7\columnwidth]{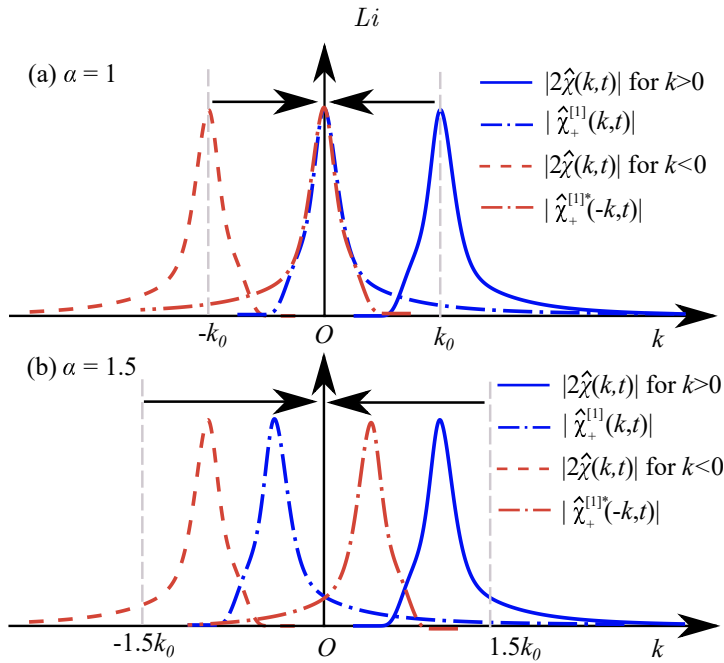}%
	\caption{Diagram of the operators in the Fourier wavenumber space in two dimensions; For the envelope transform: (a) $\alpha = 1$ and (b) $\alpha=1.5$. }
	\label{fig:operators}
\end{figure}
For later reference and simplicity,  we introduce a characteristic wave vector and angular wave frequency, denoted by $\bk_0=(k_0,0)$ and $\omega_0$, respectively, with the magnitude $k_0=|\bk_0|$. It is highlighted that the positive $x$ direction is chosen along the direction of wave vector $\bk_0$, which also accords to the main direction of wave propagation. The characteristic wave vector and frequency obey the linear dispersion relation $\omega_0=\omega(\bk_0,h)$ with $\omega(\bk,h) =\sqrt{g|\bk|\tanh |\bk|h} $. 

We introduce two operators which are demonstrated by an arbitrary temporal-spatial function, $\chi(\bx,t)$, including a Fourier transform with respect to the horizontal position vector $\bx$ and a new operator referred to as the {\it  envelope transform}, given by, respectively
\begin{linenomath}
\begin{subequations}
\begin{align} 
\hat{\chi}(\bk,t) = ~& \dfrac{1}{4\pi^2}\kint
\chi(\bx,t)\rme^{-\rmi\bk\cdot\bx}\rmd\bx,
 \label{eqs:fft_def}
\\
 \chi^{[j]}_+(\bx,t; \alpha, \beta) = ~& % \dl{\dfrac{1}{4\pi^2} }
 \rme^{\rmi j\beta \omega_0t}\kint 2\Theta[(\bk+j\alpha \bk_0)\cdot\bk_0]\hat{\chi}(\bk+\alpha\bk_0,t)
\rme^{\rmi\bk\cdot\bx}
\rmd\bk, 
 \label{eqs:op_def}
\end{align}
\end{subequations}
\end{linenomath}
in which the hat denotes a parameter transformed to the Fourier $\bk$ plane; $\alpha$, $\beta$, and $j$ are arbitrary {\it non-negative} constants which can be freely chosen to facilitate the numerical computations as will be shown in the following;  a combination of the subscript `+' and superscript `[$j$]' are used to denote the envelop transform; 
$\Theta$ denotes the Heaviside step function.  
The relationship between the two operators defined in \specialeqref{eqs:op_def}{a,b} is shown in Fig. \ref{fig:operators}.
It is seen from Fig. \ref{fig:operators},  the envelope transform is composed of three consecutive procedures.  Firstly, it collects the components of $2\hat{\chi}(\bk)$ in the wavenumber region for $\bk_0\cdot\bk>0$. Subsequently, it employs the translation operator defined as $\exp\big[-\rmi j (\alpha \bk_0\cdot\bx+\beta\omega_0 t)\big]$ in the Fourier plane. Thirdly, it operates an inverse Fourier transform given by \eqref{eqs:fft_def}. For $\alpha=0=\beta$ or $j=0$, it is understood that the envelope transform simply recovers the inverse Fourier transform with respect to $2\hat{\chi}$ in the positive wavenumber plane where $\bk\cdot\bk_0>0$. 

Due to the symmetrical properties of a Fourier transform and the definition of the envelope transform, we readily obtain
\begin{linenomath}
\begin{align}
    \chi(\bx,t) = ~&  \half  \chi\upj_+\rme^{\rmi j (\alpha \bk_0\cdot\bx-\beta\omega_0t)} + \cc,
%    \hat{\chi}(\bk,t) = ~& \hat{\chi}_+^{[j]}(\bk+j\bk_0,t)\rme^{-\rmi j\omega_0t} 
%                                        +\big[-\hat{\chi}_+^{[j]}(-\bk-j\bk_0,t)\rme^{-\rmi j\omega_0t} 
%                                         \big]^*,
\end{align}
where  $\cc$ denotes the complex conjugates,  
and
\begin{align}
	\hat{\chi}\upj_+(\bk,t)   = ~&
         2\Theta[(\bk+j\alpha \bk_0)\cdot\bk_0]\hat{\chi}(\bk+j\alpha\bk_0,t)\rme^{\rmi j\beta\omega_0t}.
	%  \left\{  \begin{array}{c}
	% 	\hat{\chi}(\bk,t)\rme^{\rmi j\beta\omega_0t},
	% 	~\text{for}~(\bk+j\alpha \bk_0)\cdot\bk_0>0, \\
	% 	0,~\text{for}~(\bk+j\alpha \bk_0)\cdot\bk_0\leq 0.
	% \end{array}\right.
%  \hat{\chi}\upj_+(\bk+j\alpha\bk_0,t)   = ~&  ~ \hat{\chi}(\bk,t)\rme^{\rmi j\beta\omega_0t} ~
%   \text{and}~
%   \hat{\chi}_+^{[j]}(\bk+j\alpha\bk_0,t)   = ~0 \notag 
\end{align}
%
% for   $(\bk+j\alpha \bk_0)\cdot\bk_0>0$ and $(\bk+j\alpha\bk_0)\cdot\bk_0\leq 0$, respectively. 
It is understood that  $\hat{\chi}_+\upj$ varies with the different combinations of the constants $(\alpha, \beta, j)$,  as shown in Fig. \ref{fig:operators} where two special cases with $\alpha=1$ and $\alpha=1.5$ are shown. 
With $j=1$, the envelope transform leads to the definition of the envelope of the wave elevation and potential at the free water surface, which will be used in \S\ref{sec:ceees}:
\begin{equation} \label{eq:ABs_def}
A(\bx,t; \alpha,\beta) = \zeta^{[1]}_+~~\text{and}~~B_{s}(\bx,t; \alpha,\beta) = \psi^{[1]}_+,
\specialnumber{a,b}
\end{equation} 
thereby,
\begin{equation} \label{eq:zt_psi_AB}
  \zeta = \half A \rme^{\rmi  (\alpha \bk_0\cdot\bx-\beta\omega_0t)} + \cc 
  ~\text{and}~
  \psi = \half B_s  \rme^{\rmi  (\alpha \bk_0\cdot\bx-\beta\omega_0t)} + \cc.
  \specialnumber{a,b}
\end{equation}
\end{linenomath}
%
% If we choice $\alpha=\beta$ in \specialeqref{eq:ABs_def}{a,b}, the definition of envelopes $A$ and $B_s$ will be the same as used in a nonlinear Schr\"{o}dinger (NLS) equation for surface water waves, e.g., [] and [] by \cite{trulsen00} and \cite{li21nls}, respectively.  
The main intention of allowing for an arbitrary choice of ($\alpha,\beta,j$) is to facilitate numerical implementations and, therefore, improve the numerical efficiency. For example, with the choice of $\alpha=0$ and $\beta=1$, we will show in \S\ref{sec:comprisons}  that the computational efficiency of a HOS method can be improved. With $\alpha=\beta=1$ and additional assumptions required, the CEEEs would be reduced to a third-order accurate NLS equation-based model which has been demonstrated with an excellent performance. 

%-------------%---%----%----%----%--------------%
%                    section                    %
%-------------%---%----%----%----%--------------%
%
\section{A review of two methods} 
\label{sec:review}
In this section, we review two methods for the description of non-breaking surface waves, which are the HOS method presented \S\ref{sec:reviewHOS} and a so-called traditional perturbation method in \S\ref{sub:tra_pertb}.  Both methods rely on unknown velocity potential being expressed in a form of perturbation expansion and Fourier transform as explained in \S\ref{sub:pertb_Ham}. Their distinctive difference lies in that  they seek different approaches for the unknowns: surface elevation, the potential and vertical velocity on the free water surface. 
\subsection{The HOS method}\label{sec:reviewHOS}
The HOS method proposes to solve the fully nonlinear (potential flow) boundary value problem in a Hamiltonian theory as introduced in \S\ref{sub:bcsf_Ham} for two main unknowns which are the surface elevation ($\zeta$) and the potential  ($\psi$) on the free water surface. It consists of two procedures. It firstly seeks to express the unknowns including the velocity potential ($\Phi$) and the vertical velocity 
({$W(\bx,t)$) 
on the free water surface in a form of the function of $\zeta$ and $\psi$, which will be presented in \S\ref{sub:hos_ss}. Secondly, through using the boundary conditions \specialeqref{eq:bcnew_sf}{a,b} it leads to the evolution equations which can be numerically solved for the two main unknowns, as presented in \S\ref{sub:hos_eqs}. 
\subsubsection{Solution structure for a finite uniform depth} \label{sub:hos_ss}
We proceed to explaining how the
 velocity potential and vertical velocity can be expressed in a form of functions of both $\zeta$ and $\psi$.  Following \cite{dommermuth87} and \cite{west87},  the HOS method proposes to letting:
\begin{linenomath}
\begin{equation}\label{eq:phi_exp}
\Phi^{(1)}(\bx,0,t) = \psi(\bx,t) \equiv \Phi(\bx,\zeta(\bx,t),t), 
\end{equation}
which, due to the perturbation expansion \eqref{eq:phizeta_pertb}, leads to the  velocity potential on the free water surface given by
\begin{align} \label{eq:phi_on_sf}
\Phi(\bx,\zeta,t)  =  \sum\limits_{m=1}^{M} \Phi^{(m)}(\bx,z,t)
~\text{for}~{z=\zeta}. 
\end{align}
An expression for $\Phi^{(m)}$ for $m>1$ can be obtained through the subsequent procedures: Taylor expanding the terms on the right hand side of \eqref{eq:phi_on_sf}  about $z=0$,   inserting \eqref{eq:phi_exp}, and collecting the same orders in wave steepness. Hence, $\Phi^{(m)}$ is expressed as functions of the lower-order parameters as follows
\begin{equation}\label{eq:phim0_def}
\Phi^{(m)}_0(\bx,t) = - \sum_{k=1}^{m-1} \dfrac{1}{k!}\zeta^k\p_z^k \Phi^{(m-k)}(\bx,z,t)~\text{for}~z=0~\text{and}~m \in \{2,3,...M\},
\end{equation}
where, as noted, the subscript `0' denotes the parameters evaluated at a still water surface, $z=0$; i.e., $\Phi^{(m)}_0 \equiv\Phi^{(m)}(\bx,0,t)$. The expression \eqref{eq:phim0_def} suggests that, if $\psi$ and $\zeta$ are given, $ \Phi^{(m)}$ for $m>1$ will be obtained in sequence from the lowest to higher orders.  Similarly, the vertical velocity ($W$) on the unknown free water surface can be obtained from Taylor expanding its definition given by \specialeqref{eq:defs_psiW}{b} about $z=0$ and inserting the expression for $\Phi^{(m)}_0$ leads to
\begin{subequations}
\label{eq:Wm_hos} 
	\begin{align} 
	W(\bx,t) = & \sum_{m=1}^{M} \eps^m W^{(m)}(\bx,t),
	~\text{with}~
	\label{eq:W_def}
	\\
	W^{(m)}(\bx,t) = & \sum_{k=0}^{m-1} \dfrac{\zeta^k}{k!}\p_z^{k+1} \Phi^{(m-k)}(\bx,z,t)~\text{for}~z=0~\text{and}~m\in  \{1,2,...,M \}. 
	\end{align}
\end{subequations}
\end{linenomath}
Therefore, $W^{(m)}$ can be obtained in sequence from the lowest order $m=1$ to higher orders due to \eqref{eq:phij_ifft} for $\Phi^{(m)}(\bx,z,t)$, \eqref{eq:phim0_def}  for $\Phi^{(m)}_0(\bx,t)$,  and \eqref{eq:Wm_hos}, with $\psi$ and $\zeta$ as input.

\subsubsection{The $M$-th order accurate equations in the HOS method}
\label{sub:hos_eqs}
Inserting the perturbed solution \eqref{eq:phi_exp} for $\psi$ and \specialeqref{eq:Wm_hos}{a,b} for $W$ into the boundary conditions \specialeqref{eq:bcnew_sf}{a,b} and keeping the terms up to order $M$ gives rise to
\begin{linenomath}
\begin{equation}\label{eq:ode_zetapsi}
\p_t \zeta  - \p_z\Phi^{(1)}  =  \mw_{M}(\bx,t) 
~\text{and}~	
\p_t\psi + g\zeta  = ~  \mt_{M}(\bx,t) 
~\text{for}~z=0,
\specialnumber{a,b}
\end{equation}
where both the super- and sub-script `$M$' denotes the truncated order of accuracy in wave steepness, 
\begin{equation}\label{eq:mw_mt}
\mw_{M}(\bx,t)  = \sum_{m=1}^M \mw^{(m)}(\bx,t)~
\text{and}~
\mt_{M}(\bx,t)  = \sum_{m=1}^M \mt^{(m)}(\bx,t),
\specialnumber{a,b}
\end{equation}
where $\mw_{1}\equiv \mw^{(1)} =0$ and $\mt_{1}\equiv \mw^{(1)}=0$, and the nonlinear forcing terms, $\mw^{(m)}(\bx,t)$ with $m\geq 2$, are non-vanishing and given by
\begin{subequations}\label{eq:mwm_hdef}
	\begin{align}
	\mw^{(2)}= ~ & W\snd
	- \nabla\psi\cdot\nabla\zeta,~\\
	\mw^{(3)} =  ~& W\trd+ W\fst (\nabla\zeta)^2,
	\\
	\mw^{(4)}= ~& W^{(4)}  + W\snd(\nabla\zeta)^2,
	~\text{and}~\\
	\mw^{(5)} =~&  W^{(5)}  + W\trd(\nabla\zeta)^2,
	\end{align}
\end{subequations}
which are explicitly expressed up to the fifth order in wave steepness; similarly, $\mt^{(m)}(\bx,t)$ are
\begin{subequations} \label{eq:mtm_hdef}
	\begin{align}
	\mt^{(2)} =~&  \half ( W\fst)^2-\half (\nabla\psi)^2,~\\
	\mt^{(3)} = ~&  W\fst W\snd , ~\\
	\mt^{(4)}= ~& \half (W\snd)^2+ W\fst W\trd+ \half  (W\fst)^2(\nabla\zeta)^2, ~\\
	\mt^{(5)} = ~& W\snd W\trd + W\fst W^{(4)} + (W\fst W\snd)(\nabla\zeta)^2.
	\specialnumber{a,b}
	\end{align}
\end{subequations}
\end{linenomath}
We can summarize that the HOS method has derived the $M-$th order accurate equations \specialeqref{eq:ode_zetapsi}{a,b} which can be numerically solved for $\zeta$ and $\psi$, with $\Phi^{(m)}$ and $W^{(m)}$ obtained from \eqref{eq:phim0_def} and \eqref{eq:Wm_hos}, respectively, in sequence from the lowest order to higher orders 
%\rep{\citep{dommermuth87,west87}}
\citep{west87}. The nonlinear terms on the right hand side of \specialeqref{eq:ode_zetapsi}{a,b} are given by \eqref{eq:mwm_hdef} and \eqref{eq:mtm_hdef}.

%-------------%---%----%----%----%--------------%
%               section                         %
%-------------%---%----%----%----%--------------%
%
\subsection{Traditional perturbation method}
\label{sub:tra_pertb}
%
% {\color{red} Should mention that the traditional perturbation method solves the BVP in sequence from the lowest order to higher orders. }
%
% Before proceeding to the details, I first note that the notations used in this section will only be used in \S\ref{sub:comp_pertb} for comparisons and completeness. 
%
I highlight that the derivations presented in this section will only be used in \S\ref{sub:comp_pertb} for comparisons and completeness.
Different from the methods based on the boundary conditions given by \specialeqref{eq:bcnew_sf}{a,b} in \S\ref{sub:bcsf_Ham}, I refer to one of these which have the following features as a traditional perturbation method.
It primarily seeks the approximate solution to the boundary value problem described by \eqref{eq:gvn_lap}, \specialeqref{eq:gvn_sf}{a,b}, and \eqref{eq:bc_sb}. The boundary conditions \specialeqref{eq:gvn_sf}{a,b} on the free water surface are especially expanded about the still water surface $z=0$ for all $z$-dependent parameters, in contrast to the boundary conditions given by 
\specialeqref{eq:ode_zetapsi}{a,b} where only the vertical velocity is expanded about $z=0$. The primary unknowns are the velocity potential on a still water surface and the surface elevation which are expressed in a form of power series in wave steepness. Substituting these approximate expressions into the boundary value problem given by \eqref{eq:gvn_lap}, \specialeqref{eq:gvn_sf}{a,b} expanded about $z=0$, and \eqref{eq:bc_sb}, and collecting the same orders in wave steepness will lead to the boundary value problems at different order in wave steepness. These boundary value problems are solved in sequence from the first to $M-$th order in wave steepness. 
A few examples which are based on a traditional perturbation method are \cite{chu71}, \S 13 by \cite{mei05}, a NLS equation-based model like \cite{davey74,dysthe179,trulsen00, slunyaev05,li21nls}, the fifth-order Stokes waves by \cite{fenton85}, and the second-order broadband framework by \cite{lili21}. 

The main derivations needed in a traditional perturbation method are explained in the following, for which the leading-order approximations are kept only to the second order in wave steepness for simplicity. In order to indicate the differences with the main results presented in most chapters of this paper, a prime is added to denote the parameters used in a traditional expansion method and  subscripts are used to denote the different orders in wave steepness and wave harmonics. An approximate form for both the unknown velocity potential and elevation are assumed, to the second order in wave steepness $\eps_0$
\begin{linenomath}
\begin{equation}\label{eq:phi_zete_prime}
    \Phi = \eps_0 \Phi'_{11}+ \eps^2_0 \underbrace{(\Phi'_{22} + \Phi'_{20})}_{\equiv \Phi'_2(\bx,z,t)}
    ~\text{and}~
    \zeta= \eps_0 \zeta'_{11}+ \eps^2_0 \underbrace{(\zeta'_{22} + \zeta'_{20})}_{\equiv \zeta'_2(\bx,t)},
    \specialnumber{a,b}
\end{equation}
where $\eps_0$ denotes the dimensionless steepness of linear waves which obeys $\mo(\eps_0)\sim\mo(k_0\zeta_{11})$ to primarily distinguish it from $\eps$ defined in \S\ref{sub:pertb_Ham}; the subscript `$mj$' denotes $\mo(\eps_0^m)$ and the $j$-th wave harmonic;  $\zeta'_{mj}=\zeta'_{mj}(\bx,t) $ and $\Phi_{mj}'= \Phi'_{mj}(\bx,z,t)$. The potential and vertical velocity on the free water surface are given in a form of Taylor expansion about $z=0$, to the second order in wave steepness $\eps_0$
\begin{equation}
    \psi = \eps_0 \Phi'_{11} +\eps^2_0  \zeta'_{11}\p_z\Phi'_{11}
    ~\text{and}~
    W=  \eps_0 \p_z\Phi'_{11} +  \eps^2_0\zeta'_{11}\p_{zz}\Phi'_{11}
    ~\text{for}~z=0.
    \specialnumber{a,b}
\end{equation}
Inserting \specialeqref{eq:phi_zete_prime}{a,b} for unknown potential and elevation, respectively, into the surface boundary conditions \specialeqref{eq:gvn_sf}{a,b}, expanding the equations about $z=0$,  collecting the terms at second order in $\eps_0$ leads to
\begin{subequations}\label{eq:bc_sf_prime}
\begin{align}
    \p_t \zeta'_{2} -\p_z\Phi'_{2} = ~& g\zeta'_{11}\p_{zz}\Phi'_{11} - \nabla\zeta'_{11}\cdot\nabla\Phi'_{11},
    \\
    \p_t\Phi'_{2} + g\zeta'_{2} = ~& \zeta'_{11}\p_{tz}\Phi'_{11}-\half (\nabla_3\Phi'_{11})^2, 
\end{align}
\end{subequations}
\end{linenomath}
which are used to solve for the unknowns (i.e. $\zeta_{2}$ and $\Phi_2$) at second order with the linear parameters obtained from the linearized equations of \eqref{eq:gvn_lap}, \specialeqref{eq:gvn_sf}{a,b}, and \eqref{eq:bc_sb}, see, e.g., \S 13 by \cite{mei05}. With the forcing terms on the right hand side of \specialeqref{eq:bc_sf_prime}{a,b} being separated according to the wave harmonics, the unknown fields with the subscript `$mj=22$' and `$mj=20$' can be obtained due to the second-order super- and sub-harmonic waves, respectively \citep{li21a}.  

A different but equivalent framework to \specialeqref{eq:bc_sf_prime}{a,b} has been proposed by \cite{lili21} where envelopes have been introduced, which are the primary unknowns for the waves of different harmonics up to the second order in wave steepness, $\eps_0$. The elevation and potential are in particular given by

\begin{subequations}\label{eq:sfbc_li21}
	\begin{linenomath}
\begin{align}
    \zeta = ~& \half \eps_0 A_{11}'\rme^{\rmi(\bk_0\cdot\bx-\rmi\omega_0t)} +\cc 
    + \eps^2_0 \left(
    \half A'_{22}\rme^{2\rmi(\bk_0\cdot\bx-\rmi\omega_0t)}
    +\half A'_{20} +\cc
    \right) ,\\
    \Phi = ~& \half \eps_0
    B_{11}'\rme^{\rmi(\bk_0\cdot\bx-\rmi\omega_0t)} +\cc + \eps^2_0 \left(
    \half B'_{22}\rme^{2\rmi(\bk_0\cdot\bx-\rmi\omega_0t)}+\half B'_{20} +\cc \right) ,
\end{align}
where the (complex) envelopes $A'_{mj}(\bx,t)$ and $B'_{mj}(\bx,z,t)$ are the main unknowns. With the second-order elevation and potential, the envelopes obey
\begin{align}
    A'_{mj}(\bx,t)=\left[\zeta'_{mj}\right]_+^{[j]}
    ~\text{and}~B'_{mj}(\bx,z,t)=\left[\Phi'_{mj}\right]_+^{[j]}
\end{align}
\end{linenomath}
for `$mj=11$', `$mj=20$', and `$mj=22$'. %\ed{Due to the  Laplace equation and the seabed boundary condition}
Due to the Laplace equation and the seabed boundary condition, the vertical structure of the envelope  $B'_{mj}$ is obtained and given by
\begin{linenomath}
\begin{align}
		B'_{mj}(\bx,z,t) = \kint \hat{B}'_{mj}(\bk,t) \dfrac{\cosh |\bk+j\bk_0|(z+h)}{\cosh |\bk+j\bk_0|h }\rme^{\rmi\kx}\rmd\bk,
\end{align}
\end{linenomath}
%%%
where $\hat{B}'_{mj}(\bk,t) $ denotes the envelope $B'_{mj}(\bx,z,t)$ at $z=0$ transformed to the Fourier space.
\end{subequations}
Similarly, inserting \specialeqref{eq:sfbc_li21}{a,b} into \specialeqref{eq:gvn_sf}{a,b}, expanding the $z$-dependent wave parameters about $z=0$, collecting the terms at the second order, and separating the wave harmonics leads to the boundary conditions for the second-order superharmonic waves on a still water surface, $z=0$
\begin{linenomath}
\begin{subequations}\label{eq:ABp_22}
\begin{align}
    (\p_{t}-2\rmi\omega_0)A'_{22}-\p_{z}B'_{22} =~&  \half g A'_{11}\p_{zz}B'_{11} - \half (\nabla+\rmi\bk_0)A'_{11}\cdot(\nabla+\rmi\bk_0)B'_{11},
    \\
     (\p_{t}-2\rmi\omega_0)B'_{22}+gA'_{22} =~&  \half A'_{11}\p_{tz}B'_{11} - \half [(\nabla+\rmi\bk_0)B'_{11}]^2
     - \half (\p_zB'_{11})^2;
\end{align}
\end{subequations}
the boundary conditions for the second-order subharmonic waves on a still water surface
\begin{subequations}\label{eq:ABp_20}
\begin{align}
    \p_tA'_{20}-\p_{z}B'_{20} =~&  \left[ \mR\left( \half g A^{'*}_{11}\p_{zz}B'_{22} - \half (\nabla-\rmi\bk_0)A^{'*}_{11}\cdot(\nabla+\rmi\bk_0)B'_{11}\right) \right]^{(0)}_{+},
    \\
    \p_t B'_{20}+gA'_{20} =~& \left[ \mR\left( \half A^{'*}_{11}\p_{tz}B'_{22} - \half |(\nabla+\rmi\bk_0)B'_{11}|^2
     - \half |\p_zB'_{11}|^2\right) \right]^{(0)}_{+},
\end{align}
\end{subequations}
\end{linenomath}
where $\mR$ denotes the real component.
With linear envelope $A'_{11}$ and $B'_{11}$ solved from the linearized problem, the second-order envelopes can be obtained from \specialeqref{eq:ABp_22}{a,b} and \specialeqref{eq:ABp_20}{a,b} for the super- and sub-harmonic waves, respectively. 

Both the envelope equations, \specialeqref{eq:ABp_22}{a,b} and \specialeqref{eq:ABp_20}{a,b}, for the envelopes as well as the equations, \specialeqref{eq:bc_sf_prime}{a,b}, for the second-order elevation and potential will be used in \S\ref{sub:comp_pertb} for a comparison with the CEEEs derived in \S\ref{sec:ceees}, showing how the CEEEs can recover to these equations if waves are  considered only up to the second order in wave steepness, $\eps_0$. 

%-------------%---%----%----%----%--------------%
%                    section                    %
%-------------%---%----%----%----%--------------%
%
\section{Coupled envelope evolution equations in a Hamiltonian theory} \label{sec:ceees}
In this section,  we derive the new equations, referred to as the {\it  coupled envelope evolution equations} (CEEEs), in a Hamiltonian theory for numerical solutions, which are the main results of this paper. To this end, the starting point is a new pair of canonical variables shown in \S\ref{sub:newVar}. In contrast to the HOS, the new canonical variables become our primary unknowns which can be numerically solved for. Especially the wave parameters (velocity potential and vertical velocity) at different orders in wave steepness will be obtained through  harmonic separations, as shown in \S\ref{sub:3methods} and \S\ref{sub:sep_harm}. The CEEEs describing the evolution of the new canonical variables are presented in \S\ref{sub:ceees}. Similar to the HOS method, the CEEEs can be derived up to an arbitrary order in wave steepness, with the general expressions presented in Appendix \ref{app:Morder}. 
%
%---%---%----%----%----%
%              section             %
%---%---%----%----%----%
%
\subsection{A new pair of canonical variables}\label{sub:newVar}
It is understood that $\zeta$ and $\psi$ is a pair of canonical variables \citep{zakharov68,krasitskii94}. Due to the definition of the envelopes, we proceed to show that 
$A$ and $B_s$ are a new pair of  canonical variables. It is understood that, due to the properties of Fourier transform,  the following identities hold
\begin{linenomath}
\begin{equation}
    \hzt(\bk) = \hzt^*(-\bk),~ \hzt(-\bk) = \hzt^*(\bk),~
     \hat{\psi}(\bk) = \hat{\psi}^*(-\bk),~\text{and}~ \hat{\psi}(-\bk) =  \hat{\psi}^*(\bk),
\end{equation}
where the asterisk `*' denotes the complex conjugates. Let $H'$ be the Hamiltonian for $\hzt$ and $\hat{\psi}$, suggesting that
\begin{equation}\label{eq:A_Horig}
    \p_t\hzt = \dfrac{\delta H'}{\delta \hat{\psi}^*}
    ~\text{and} ~
    \p_t\hat{\psi}= - \dfrac{\delta H'}{\delta \hzt^*},
    \specialnumber{a,b}
\end{equation}
where the Hamiltonian $H'=H'(\hzt,\hzt^*,\hat{\psi}, \hat{\psi}^*)$ and $\delta$ denotes the functional derivative. By definition we obtain
$\hzt(\bk+\alpha\bk_0) = \hat{A}(\bk,t)\exp{(-\rmi\beta\omega_0t)}$ and $\hat{\psi}(\bk+\alpha\bk_0)  = \hat{B}(\bk,t)\exp{(-\rmi\beta\omega_0t)}$, inserting which into the Hamiltonian $H'$, we obtain 
%%
%\begin{align}
%    L = \rme^{-\rmi\omega_0t}( \p_{t} \hat{A}-\rmi \omega_0  \hat{A})~\text{for}~(\bk+\bk_0)\cdot\mathbf{e}_x>0 ,
%\end{align}
%%
%If we write, for instance, 
%%
\begin{align}\label{eq:HHprim}
     H'(\hzt,\hat{\psi},\hzt^*, \hat{\psi}^*) \equiv H(\hat{A},\hat{B}_s,\hat{A}^*,\hat{B}_s^*),
\end{align}
where $H$ denotes the new Hamiltonian obtained through replacing the elevation and potential with their envelopes in $H'$. Next, we perform the following functional derivatives based on \eqref{eq:HHprim} and obtain
\begin{equation}
    \delta_{\hat{\psi}^*} H' =  %   \dfrac{\mathrm{D} B^*}{\mathrm{D} \psi^*} \delta_{B^*} H'    =
             \exp{(-\rmi\beta\omega_0t)} \delta_{B^*} H
    ~\text{and}~
    \delta_{\hzt^*} H' =  \exp{(-\rmi\beta\omega_0t)} \delta_{A^*} H, 
    \specialnumber{a,b}
\end{equation}
inserting which into the right hand side of \specialeqref{eq:A_Horig}{a,b}, replacing the elevation and potential with their envelopes on the left hand side of \specialeqref{eq:A_Horig}{a,b}, and eliminating the factor $\exp{(-\rmi\beta\omega_0t)}$ gives rise to 
\begin{equation} \label{eq:AhAB}
   \p_t \hat{A} -\rmi\omega_0 \hat{A} = \delta_{B_s^*} H
  ~ \text{and}~
 -\p_t \hat{B}_s +\rmi\omega_0 \hat{B}_s = \delta_{A^*} H.
 \specialnumber{a,b}
\end{equation}
Multiplying \specialeqref{eq:AhAB}{a,b} by $\hat{B}_s^*$ and $\hat{A}^*$, respectively, leads to
\begin{equation}
    \hat{B}^*\p_t \hat{A} -\rmi\omega_0 \hat{B}_s^*\hat{A} =   \hat{B}_s^*\delta_{B^*_s} H'
     ~ \text{and}~
   - \hat{A} ^*\p_t \hat{B}_s +\rmi\omega_0 \hat{A} ^* \hat{B}_s =\hat{A} ^*\delta_{A^*} H',
   \specialnumber{a,b}
\end{equation}
based on which we introduce a new Hamiltonian defined as
\begin{align}
   H_{AB}(\hat{A},\hat{B}_s,\hat{A}^*,\hat{B}_s^*)= \int  H' + \rmi \omega_0 \left[ \hat{B}_s^*\hat{A}  -\hat{A} ^* \hat{B}_s\right]\rmd \bk.
\end{align}
Performing the following functional derivatives on the new Hamiltonian $H_{AB}$ leads to
\begin{equation}\label{eq:AdHnew}
   \delta_{B_s^*} H_{AB} = \delta_{\hat{B}_s^*} H' + \rmi \omega_0 \hat{A}
   ~ \text{and}~
  \delta_{A^*} H_{AB} =  \delta_{\hat{A}^*} H' - \rmi \omega_0 \hat{B}_s.
  \specialnumber{a,b}
\end{equation}
Inserting \specialeqref{eq:AdHnew}{a,b} for $ \delta_{B^*} H'$ and $ \delta_{A^*} H'$ into the right hand side of \specialeqref{eq:AhAB}{a,b}, respectively, leads to
\begin{equation}\label{eqs:ham_hAB}
     \p_t\hat{A} =  \dfrac{\delta H_{AB}'}{\delta \hat{B}_s^*}
     ~ \text{and}~
      \p_t\hat{B}_s =-\dfrac{\delta H_{AB}'}{\delta \hat{A}^*},
      \specialnumber{a,b}
\end{equation}
\end{linenomath}
meaning that $\hat{A}$ and $\hat{B}_s$ are a pair of canonical variables. As noted in \cite{krasitskii94}, due to that the inverse Fourier transform is a canonical one and therefore, equations \specialeqref{eqs:ham_hAB}{a,b} also imply that envelope $A$ and $B_s$ are  a pair of canonical variables. 

\subsection{Three different methods for the evaluation of a quadratic term}
\label{sub:3methods}
In this paper we take advantage of the symmetrical properties of  Fourier transform.  The separation of wave harmonics presented in \S \ref{sub:sep_harm} builds upon two key features. Firstly, it is understood that nonlinear terms (e.g., $\mw^{(3)}$ and $\mt^{(4)}$) at orders higher than the second can always be written in a form of the linear superposition of the product of two parameters.  Using $\mt^{(4)}$ described by \specialeqref{eq:mtm_hdef}{c} as an example.  We define
\begin{linenomath}
\begin{align}
   W_{sq}\fst = (W\snd)^2 ~~
   \text{and}~~
   \zeta_x^{(2)} = (\nabla\zeta)^2, 
\end{align}
inserting which into \specialeqref{eq:mtm_hdef}{c} leads to 
\begin{align}
   	\mt_0^{(4)}= ~& \half (W\snd)^2+ W\fst W\trd+ \half  W_{sq}\fst  \zeta_x^{(2)},
\end{align}
which is in a form of the linear superposition of quadratic terms but it is obvious that it is not at second order in wave steepness.  Second,  by virtue of the symmetrical properties of Fourier transform, we next consider a function of two arbitrary real parameters $\chi(\bx)$ and $\xi(\bx)$ defined as  $\mathbf{f}(\bx) = \nabla\chi(\bx)\xi(\bx)$. Assuming the Fourier transform of both $\chi(\bx)$ and $\xi(\bx)$  exist, $\mathbf{f}(\bx)$ can also be expressed in a form of inverse Fourier transform
\begin{subequations}
	\begin{align}
   \mathbf{f}(\bx)= ~& 
%   \dl{\dfrac{1}{16\pi^4}}
   \kint  \rmi\bk_1\hat{\chi}_1\hat{\xi}_2\rme^{\rmi(\bk_1+\bk_2)\cdot\bx}\rmd \bk_1\rmd \bk_2
	\label{eq:chixi_fft}
	~\text{or}~
	\\
	\mathbf{f}(\bx) = ~& \left[
%  \dl{\dfrac{1}{4\pi^2}} 
  \kint \rmi\bk_1 \hat{\chi}_1\rme^{\rmi\bk_2\cdot\bx}\rmd \bk_1
	\right]\times 
 \left[
%\dl{ \dfrac{1}{4\pi^2}}
\kint  \hat{\xi}_2\rme^{\rmi\bk_2\cdot\bx}\rmd \bk_2
	\right],
	\label{eq:chixi_pseudo}
	\end{align}
\end{subequations}
where $\hat{\chi}_1 = \hat{\chi}(\bk_1)$ and $\hat{\xi}_2=\hat{\xi}(\bk_2)$.
Due to the symmetrical property of a Fourier transform (i.e., $\hat{\chi}(-\bk)= \hat{\chi}^* (\bk)$) and decomposing the entire integral region of the integrals in \eqref{eq:chixi_fft} into four equal quarters  leads to
\begin{align}
  \mathbf{f}(\bx) =  ~&  
  %\dl{\dfrac{1}{16\pi^4}}
                         \left[ \efact 
                         \int\limits_{\Gamma_{1+}}  \int\limits_{\Gamma_{2+}}
                       \rmi\bk_1\hat{\chi}_1\hat{\xi}_2 
                         \rme^{\rmi(\bk_1-\alpha\bk_0+\bk_2-\alpha\bk_0)\cdot\bx+2\rmi\beta\omega_0t} 
                       \rmd\bk_1\rmd \bk_2  + \cc \right]
                      \notag\\
             +    ~&   
   %          \dl{\dfrac{1}{16\pi^4}}
                 \left[
                 \int\limits_{\Gamma_{1+}}  \int\limits_{\Gamma_{2+}}  \rmi\bk_1
                 \hat{\chi}_1\hat{\xi}^*_2 \rme^{\rmi(\bk_1-\alpha\bk_0)\cdot\bx -\rmi (\bk_2-\alpha\bk_0)\cdot\bx}               
                 \rmd\bk_1\rmd \bk_2  
                 + \cc  \right],
                 \label{eq:chixi_fft+}
\end{align}
in which $\Gamma_{j+}$ for $j=1$ and $j=2$ defines the region where $\bk_j\cdot\bk_0> 0$.  Replacing the terms which correspond to the definition of the envelope transform in \eqref{eq:chixi_fft+} leads to
\begin{align}\label{eq:chixi_new}
 \mathbf{f}(\bx)= \left[\dfrac{1}{4} (\nabla+\rmi\alpha\bk_0) \chi^{[1]}_+{\xi}^{[1]}_+ \efact + \cc \right] + 
                                        \left[ \quard (\nabla+\rmi\alpha\bk_0)\chi^{[1]}_+ \big(\xi^{[1]}_+\big)^*+ \cc\right].
\end{align}
\end{linenomath}
The above discussion suggests that the evaluation of $ \mathbf{f}(\bx)$ admits at least three different forms, which lead to the main differences between different methods for the description of water waves as explained in the following. The HOS method relies on \eqref{eq:chixi_pseudo} which is typically used when the derivatives with respect to $x$ or $y$ of a parameter are involved based on a pseudo-spectral method. The Hasselmann/Zakharov integral equation \citep{hasselmann62,zakharov68,stiassnie84b,krasitskii94} uses \eqref{eq:chixi_fft} for the evaluation of $\mathbf{f}(\bx)$. The CEEEs proposed in this paper rely on \eqref{eq:chixi_new} instead, which is in principle the so-called separation of wave harmonics and does not rely on the narrowband assumption. In contrast to the HOS and  Zakharov equations, we will obtain the equations for unknown envelopes in the following sections for the CEEEs  in a manner similar to a NLS-based model but at no cost of the accuracy.  

\subsection{Separation of  wave harmonics}
\label{sub:sep_harm}
We proceed to seek a different but equating expression for the description of wave fields, especially the potential and vertical velocity, in a form of functions of unknown envelope $A(\bx,t)$ and $B_s(\bx,t)$ through the separation of harmonics presented in  \S \ref{sub:3methods}. Doing so will permit us to take advantage of a pseudo-spectral Fourier method in a more coarse and larger grid at a little additional cost of numerical computations while not at the expenses of the accuracy, compared with the HOS method.  Based on the solution structure for both the potential and velocity presented in \S~\ref{sub:hos_ss}, we start with \S~\ref{sub:cees_m12} for the velocity potential ($\Phi^{(m)}$) in a form of function of $A$ and $B_s$ up to second order in wave steepness. % \dl{The extension to  an arbitrary order can in principle be derived.} 
For simplicity, an example of $M$ up to $M=4$ for both the potential and vertical velocity is presented in \S~\ref{sub:cees_orderM} with the forcing terms derived in \S~\ref{sub:cees_wt}.The general procedures for the derivations up to an arbitrary order are presented in Appendix \ref{app:Morder}. The new framework presented in this section can be made numerically feasible with a varying parameter of $M$, similar to the HOS method.

\subsubsection{Methodology illustration}
\label{sub:cees_m12}
We proceed to explain the fundamental methodology of the new envelope framework using the velocity potential in the first and second-order approximations as examples. % 
% \rep{}{Velocity}
%\ed{\section{Velocity}}
%
For later reference, we define
\begin{linenomath}
\begin{equation}
   B_0(\bx,t) = B_s(\bx,t)~
   \text{and}
   ~B^{(11)}(\bx,z,t) = B(\bx,z,t),
   \specialnumber{a,b}
\end{equation}
%
% where we note that the superscript in a form of `($mj$)' denotes $\mo(\eps^m)$ and the $j$-th wave harmonic.  
Due to the definitions $\Phi_0\fst=\psi$ and \specialeqref{eq:zt_psi_AB}{b} also for the potential on the free water surface, we propose to let $B(\bx,z,t)=\big[\Phi\fst(\bx,z,t) \big]^{[1]}_+$ and therefore
\begin{subequations}%
\begin{align} \label{eq:Phi11_def}
    \Phi\fst(\bx,z,t) = ~\half B(\bx,z,t)\efac + \cc, 
\end{align}
where the relation $\hat{B}_0(\bk,t) = \hat{B}_s(\bk,t) $ holds by definition, and the Laplace equation and seabed boundary condition require
\begin{align}\label{eq:B11_def}
B(\bx,z,t) =~  
 \kint \hat{B}_0(\bk,t)  \dfrac{\cosh |\bk+\alpha \bk_0|(z+h)}{\cosh |\bk+\alpha\bk_0|h}
\rme^{\rmi\bk\cdot\bx}
\rmd\bk. 
\end{align}
\end{subequations}
The expression \eqref{eq:Phi11_def} for $\Phi\fst$ is in a new form we intend to obtain at the first order in wave steepness. We next proceed to $m=2$ for the velocity potential $\Phi\snd$.  Inserting  \specialeqref{eq:zt_psi_AB}{a,b} and \eqref{eq:Phi11_def}  into \eqref{eq:phim0_def} for $\Phi^{(2)}$, we obtain,
\begin{equation}\label{eq:phi2_sep}
   \Phi_0\snd = \Phi\sz_0(\bx,t) + \Phi_0\ssd(\bx,t)~
   \text{,and thus}  ~
    \Phi\snd = \Phi\sz(\bx,z,t) + \Phi\ssd(\bx,z,t),
    \specialnumber{a,b}
\end{equation}
where the superscript in a form of `($mj$)' denotes $\mo(\eps^m)$ and the $j$-th wave harmonic. Especially the superscripts `$20$' and `$22$'  denote the potential for the second-order subharmonic and superharmonic bound waves, respectively, which are obtained through the separation of harmonics given by
\begin{equation}\label{eq:phi2j_0}
{\Phi}\sz_0=  -\quard A^*\p_zB+\cc
~\text{and}~
\Phi\ssd_0 = - \quard A\p_zB\rme^{2\rmi (\alpha\bk_0\cdot\bx-\beta\omega_0t)} +\cc.
~\text{for}~z=0.
\specialnumber{a,b}
\end{equation}
We next define the envelope of the two potentials as
\begin{equation}\label{eq:B2j_idea}
   B^{(2j)}(\bx,z,t) =\big[ \Phi^{(2j)}\big]^{[j]}_{+}~~
   ~\text{and}~      
   B^{(2j)}_0(\bx,t) = \big[ \Phi^{(2j)}_0\big]^{[j]}_{+}~~
   \text{for}~j=0~~\text{and}~~j=2
   \specialnumber{a,b}
\end{equation}
which show the relations between the envelope and potential due to the second-order super- ($j=2$) and sub-harmonic ($j=0$) waves. As the second-order sub-harmonic envelopes based on \specialeqref{eq:B2j_idea}{a,b} and \specialeqref{eq:phi2j_0}{a} depend only on slowly varying envelopes $A$ and $B$, they are used in the new framework. Nevertheless, the envelopes of the 
waves with second harmonic given by \specialeqref{eq:B2j_idea}{a,b} and \specialeqref{eq:phi2j_0}{b} have the same temporal and  spatial variation as the second-order velocity potential used in the HOS method, and thereby do not introduce merits in the efficiency in numerical implementations compared with using the HOS method. For making a difference, the second-order superharmonic envelope in the form as follows is used instead
\begin{align}
    B\ssd_0 = -\half A\p_zB
   ~\text{for}~ z=0,
\end{align}
which  depends only on the slowly varying envelopes. %
Due to the definition of the second-order envelopes $B^{(2j)}$,  the second-order potentials can also be given by
\begin{align}\label{eq:cees_phi2j}
     \Phi^{(2j)} = \half B^{(2j)}(\bx,z,t)\rme^{\rmi j(\bk_0\cdot\bx-\omega_0t)} +\cc
     ~\text{for}~
     j=0~\text{and}~j=2.
\end{align}
The Laplace equation for $\Phi^{(2j)} $ and the seabed boundary condition lead to
\begin{align}
    B^{(2j)}(\bx,z,t) =~& % \dl{\dfrac{1}{4\pi^2}} 
    \kint \hat{B}^{(2j)}_0(\bk,t)  \dfrac{\cosh |\bk+j\alpha\bk_0|(z+h)}{\cosh |\bk+j\alpha\bk_0|h}
    \rme^{\rmi\bk\cdot\bx}
    \rmd\bk. 
\end{align}
\end{linenomath}
where $j=0$ and $j=2$ for the sub- and super-harmonic envelopes, respectively; as noted, the subscript `0'  denotes the evaluation at $z=0$ and the hat added denotes the Fourier transform. 

\subsubsection{Velocity potential and vertical velocity}
\label{sub:cees_orderM}
At second order, we have obtained an equating form for the second order potential %\dl{ expressed as \specialeqref{eq:phi2_sep}{a,b},} 
which is in a form of the linear superposition of different wave harmonics and envelopes. All second-order envelopes are functions of slowly varying envelopes in the lower order in wave steepness. Following the methodology presented in \S \ref{sub:cees_m12}, the envelope of an individual field can be obtained to arbitrary order in wave steepness, which has been given explicitly here up to the fourth order and the general expressions up to  arbitrary order are derived in Appendix \ref{app:Morder}. In particular,  we propose to obtain a new expression  for the potential at different orders in wave steepness based on the Laplace equation for an individual potential, the seabed condition, and the perturbation expansion \eqref{eq:phim0_def}. They have a general form as follows
\begin{linenomath}
\begin{equation} \label{eq:cees_Phim}
 \Phi^{(m)} = ~  \sum_{j=0}^{j=m} \Phi^{(mj)}(\bx,z,t)
 ~\text{with}~
 \Phi^{(mj)} \equiv~ \half B^{(mj)}(\bx,z,t)\efacj+ \cc, 
    \specialnumber{a,b}
\end{equation}
and therefore
\begin{subequations}\label{eq:cees_Phim0}
\begin{align} 
\Phi^{(m)}_0 = ~& \sum_{j=0}^{j=m} \Phi^{(mj)}_0(\bx,t)
~\text{with}~\\
\Phi^{(mj)}_0 \equiv ~& \half B^{(mj)}_0(\bx,t)\efacj + \cc 
\\
\equiv ~& \half \bar{\Phi}^{(mj)}_0(\bx,t)\efacj+ \cc, 
% \specialnumber{a,b}
%
\end{align}
\end{subequations}
where the velocity potential of $j$-th harmonic in the $m-$th order in wave steepness in a form as \specialeqref{eq:cees_Phim0}{b} are used in the new framework. The differences between the envelope $B_0\mj$, and $\bar{\Phi}_0\mj$ lie in that the latter is obtained based on \eqref{eq:phim0_def} which corresponds to the factor in front of the $j$-th harmonic due to $\exp[\rmi j(\alpha\bk_0\cdot\bx-\beta\omega_0t)]$, and thereby $B_0\mj=\bar{\Phi}_0\mj$ only if $m=j$. The envelopes are
\begin{equation}\label{eq:Bmj_Phimj}
   B^{(mj)}(\bx,z,t) = \big(\Phi^{(mj)}\big)_+^{[j]} 
   ~\text{and}~
    B_0^{(mj)}(\bx,t) = \big(\Phi^{(mj)}_0\big)_+^{[j]},
   \specialnumber{a,b}
\end{equation}
which hold by definition. Due to the Laplace equation and the seabed condition, we arrive at
\begin{equation} \label{eq:Bmj}
B^{(mj)}(\bx,z,t) =~  \kint \hat{B}^{(mj)}_0(\bk,t)  \dfrac{\cosh |\bk+j\alpha\bk_0|(z+h)}{\cosh |\bk+j\alpha\bk_0|h}
\rme^{\rmi\bk\cdot\bx}
\rmd\bk.
\end{equation}
Thereby, if $B_0^{(mj)}(\bx,t)$ is given, $\Phi^{(m)}$ and  $\Phi\mj$ will be explicitly obtained from \specialeqref{eq:cees_Phim}{a,b}, respectively. 
We next only have to explain how to obtain the envelope on the still water surface, $B^{(mj)}_0(\bx,t)$, and their Fourier transform having appeared in the integrand of the integral given by \eqref{eq:Bmj} in practice, following the same procedures as for $m=2$ presented in \S~\ref{sub:cees_m12}. 
Inserting \specialeqref{eq:zt_psi_AB}{a} and  \specialeqref{eq:zt_psi_AB}{b} for the surface elevation and the potential on the free water surface, respectively, into \eqref{eq:phim0_def},  we obtain in sequence up to the fourth order in wave steepness
\begin{subequations} \label{eq:Phiij}
\begin{align}
    \Phi\snd_0(\bx,t)    = ~& \Phi\ssd_0(\bx,t)  + \Phi\sz_0(\bx,t)  , \\
    \Phi\trd_0(\bx,t)      = ~& \Phi\trdf_0 (\bx,t)   +  \Phi\trdt_0 (\bx,t),
    \\
    \Phi^{(4)}_0 (\bx,t)     = ~& \Phi^{(40)}_0 (\bx,t)   +  \Phi^{(42)}_0 (\bx,t) + \Phi_0 ^{(44)}(\bx,t) ,
\end{align}
\end{subequations} 
where $\Phi^{(mj)}_0$ is the (real) potential of the $j$-th harmonic at $\mo(\eps^m)$ and 
\begin{align}
    {\Phi}_0^{(mj)}    = &  0~\text{for}~
    mj\in 
    \{21,~30,~32,~41,43
    \}.  
\end{align}
\end{linenomath}
The new framework aims to express the  non-vanishing potentials $\Phi^{(mj)}_0$  in a form as \specialeqref{eq:cees_Phim0}{b}, relying on the middle step for $\Phi^{(mj)}_0$ given by \specialeqref{eq:cees_Phim0}{c} which depends on the explicit expression for $\bar{\Phi}_0\mj$. Thereby, as noted, $\bar{\Phi}_0\mj$ are obtained from \eqref{eq:phim0_def} through collecting the $j-$th harmonics at an individual order in wave steepness from the lowest to higher orders in sequence; explicitly for $z=0$,
\begin{linenomath}
\begin{subequations}
\begin{align}
	\bar{\Phi}\sz_0=  & - \half A^*\p_zB,
	\\
        \bar{\Phi}_0\ssd =~&
        -\half A\p_zB,~ \\
	\bar{\Phi}\trdf_0= & -
	\half\left(
	\p_zB\ssd A^* + 2\mR(\p_zB)\sz A + \half |A|^2 \p_{zz}B+\quard A^2\p_{zz}  B ^*
	\right),
	\\
	\bar{\Phi}_0\trdt= ~& 
        -\dfrac{1}{8} A^2 \p_{zz}B- \half A\p_z B\ssd,~\\
	\bar{\Phi}^{(40)}_0= &  -    
        \half \left(
	\p_zB\trdf  A^*
		+ 
  {\half}\big(\p_{zz}B\sz\big) |A|^2
	+ \dfrac{1}{4}\p_{zz}B\ssd (A^2)^*
	+ \dfrac{1}{8} |A|^2A^*\p_{zzz}B
	\right),
	\\
	\bar{\Phi}^{(42)}_0= & -    \half 
	\p_zB\trdf  A
	-    \half 	\p_zB\trdt  A^*
	- \dfrac{1}{4}\p_{zz}B\ssd |A|^2
	-  \dfrac{1}{4}\mR\big(\p_{zz}B\sz\big) A^2 \notag\\
~& 	- \dfrac{1}{16} |A|^2A\p_{zzz}B - \dfrac{1}{48}A^3\p_{zzz}B^*,
         \\
	\bar{\Phi}_0^{(44)} =~& 
        -\dfrac{1}{48} A^3 \p_{zzz}B-\dfrac{1}{8}  A^2\p_{zz} B\ssd 
	                         -\half A^2\p_{z}B\trdt,
\end{align}
\end{subequations}
\end{linenomath}
where it is clear that $\bar{\Phi}\mj_0$ depends only on the slowly varying envelopes. The envelopes of the velocity potential $B\mj$, rely on their values at still water surface, $B\mj_0$, due to their explicit form given by \eqref{eq:Bmj_Phimj}. To this end, the envelopes at a still water surface $B\mj_0$, are obtained through their relation with $\bar{\Phi}_0\mj$ due to \specialeqref{eq:cees_Phim0}{b,c}.  
As noted, $B_0\mj=\bar{\Phi}_0\mj$ for $m=j$ leads to
\begin{linenomath}
\begin{subequations}\label{eqs:Bmm}
	\begin{align}
	B_0\ssd =~& 
       \bar{\Phi}_0\ssd\equiv
        -\half A\p_zB,~ \\
	B_0\trdt= ~& 
        \bar{\Phi}_0\trdt\equiv
       -\dfrac{1}{8} A^2 \p_{zz}B- \half A\p_z B\ssd ,~\\
	B_0^{(44)} =~& 
       \bar{\Phi}_0^{(44)}\equiv
        -\dfrac{1}{48} A^3 \p_{zzz}B-\dfrac{1}{8}  A^2\p_{zz} B\ssd 
	                         -\half A^2\p_{z}B\trdt.
	\end{align}
\end{subequations}
\end{linenomath}
The other non-vanishing $B_0^{(mj)} $ are 
obtained through combining the relation with $\bar{\Phi}_0\mj$ given by \specialeqref{eq:cees_Phim0}{b,c} and their definitions by \specialeqref{eq:Bmj_Phimj}{b}. Using in addition the properties of Fourier transforms, they can be especially obtained through an inverse Fourier transform as follows
\begin{linenomath}
\begin{align}\label{eqs:hBmj}
      \hat{B}^{(mj)}_0(\bk+j\alpha\bk_0,t)\etmj
       = ~&
        \Theta[(\bk+j\alpha\bk_0)\cdot\bk_0]
	\left\{ \hat{\bar{\Phi}}^{(mj)}(\bk+j \alpha\bk_0,t)\etmj
       + 
        \right.
	\notag\\
	&  
      \left.\big[\hat{\bar{\Phi}}^{(mj)}(-\bk-j\alpha\bk_0,t)\etmj
	\big]^* \right\},
	% \label{eq:hB_hbphi}
	%
% \hat{B}^{(mj)}_0(\bk+j\alpha\bk_0,t)\rme^{-\rmi j\omega_0t}  = ~& 0~ \text{for} ~(\bk+j\alpha\bk_0)\cdot\bk_0\leq 0,
\end{align}
\end{linenomath}
where $\hat{\bar{\Phi}}_0^{(mj)}(\bk,t)$ is  the Fourier transform of $\bar{\Phi}_0^{(mj)}(\bx,t)$. 
It should be highlighted that the use of \specialeqref{eqs:Bmm}{a--c}  and \eqref{eqs:hBmj} for the envelopes, $B^{(mm)}$  and $\hat{B}\mj$ with $m\neq j$, respectively, contributes to the improvement of the computational efficiency, compared with using the original definition of the envelopes given by \specialeqref{eq:Bmj_Phimj}{b}. It is by virtue of that $\bar{\Phi}\mj$ always have the same spatial (long) scale as envelope $A$ and $B_s$. However, due to the linear translation operator indicated by the independent variable $\bk+j\alpha \bk_0$  in \eqref{eqs:hBmj}, a great care in the numerical implementation would be needed with the use of a (inverse) fast Fourier transform.    
It is now understood that potential $\Phi_0^{(m)}$ at a nonlinear order in wave steepness admits three equating forms, which are in a form given by \eqref{eq:phim0_def} and 
\begin{linenomath}
\begin{align}
   \Phi^{(m)}_0
   \equiv 
   \sum_{j=0}^{m} 
   \left[
      \half B^{(mj)}_0 \efacj+ \cc
                       \right] 
   \equiv \sum_{j=0}^{m} \left[
                       \half \bar{\Phi}^{(mj)}_0 \efacj+ \cc
                       \right].
\end{align}

Inserting the expression of $\Phi^{(m)}$ and $\Phi_0^{(m)}$  given by \specialeqref{eq:cees_Phim}{a} and \specialeqref{eq:cees_Phim0}{a}, respectively,  into \eqref{eq:Wm_hos} leads to the vertical velocity given by
\begin{subequations} \label{eq:wdefs}
\begin{align}
  w\fst = \half \p_zB\efac +\cc, ~~W\fst = w\fst_0, 
\end{align}
and 
\begin{align}
  W^{(m)} =~&  \sum_{j=0}^{j=m} \left[\half\bar{W}^{(mj)}(\bx,t) \efacj + \cc \right]
 ~\text{for}~m=2,3,...
  \\
  w^{(m)} =~&  \sum_{j=0}^{j=m} \half \p_z{B}^{(mj)}(\bx,z,t) \efacj + \cc
   ~ \text{for}~m=2,3,...
\end{align}
where the non-vanishing terms are expressed as
\begin{align}
    \bar{W}^{(20)} =~& \p_zB^{(20)} + \half A^*\p_{zz}B,\\
    \bar{W}^{(22)} =~& \p_zB^{(22)} + \half A\p_{zz}B,\\ 
     \bar{W}^{(31)} =~& \p_zB^{(31)} + 
                 \half A^*\p_{zz}B\ssd + 
                   A{\mR\big(\p_{zz}B\sz\big)} +
                 \dfrac{1}{8} A^2\p_{zzz}B^* + \dfrac{1}{4} |A|^2\p_{zzz}B  ,
     \\ 
    \bar{W}^{(33)} =~& \p_zB^{(33)} + \half A\p_{zz}B\ssd + \dfrac{1}{8} A^2\p_{zzz}B,
    \\
    \bar{W}^{(40)} =~&  \p_zB^{(40)} + 
                         \half A^*\p_{zz}B\trdf + 
                           \dfrac{1}{{4}}
                           |A|^2
                           \big(\p_{zzz}B\sz\big)
                           +\dfrac{1}{8}(A^*)^2\p_{zzz}B\ssd
                           \notag\\
                          &        + \dfrac{1}{16} |A|^2 A^*\p_{zzzz}B,
    \\
    \bar{W}^{(42)} =~&  \p_zB^{(42)} + 
     \half \p_{zz}B\trdf  A
     +\half \p_{zz}B\trdt  A^*
    + \dfrac{1}{4}|A|^2\p_{zzz}B\ssd 
    +  \dfrac{1}{4}A^2\mR\big(\p_{zzz}B\sz\big) 
    \notag\\
    & + \dfrac{1}{16} |A|^2A\p_{zzzz}B
    + \dfrac{1}{{48}} A^3\p_{zzzz}B^* ,
    \\
    \bar{W}^{(44)} =~& \p_zB^{(44)} + \half A\p_{zz}B\trdt + \dfrac{1}{8} A^2\p_{zzz}B\ssd + \dfrac{1}{48}A^3\p_{zzzz}B,
\end{align}
\end{subequations}
\end{linenomath}
for $z=0$, which depend only on the {slowly} varying envelopes.

\subsubsection{The nonlinear forcing terms on a still water surface}
\label{sub:cees_wt}
%
% {\color{red} The expressions of the forcing terms in this section need to be verified numerically!!!!!!!!!!!!!!!!!!}

%
The forcing terms at different orders  in wave steepness, described by \specialeqref{eq:mwm_hdef}{a--e}  and \specialeqref{eq:mtm_hdef}{a--e}  can also be expressed in a form of function of envelope $A$ and $B_s$ (bearing in mind that $B_s=B_0$).  Based on \specialeqref{eq:mwm_hdef}{a--e}  and \specialeqref{eq:mtm_hdef}{a--e}, we obtain
\begin{linenomath}
\begin{subequations}\label{eq:mwt_defs}
\begin{align}
    \mw^{(m)}(\bx,t) = ~&\sum_{j=0}^{j=m} ~
    \underbrace{
    	\left[
    \half \bmw^{(mj)}(\bx,t)
    \efacj +\cc
         \right]
                }_{\equiv \mw^{(mj)}(\bx,t)}
    ~\text{and}~
    \\
     \mt^{(m)}(\bx,t) = ~&\sum_{j=0}^{j=m}  ~
                   \underbrace{
                 	\left[
                 	\half  \bmt^{(mj)}(\bx,t)
                      \efacj +\cc
                \right]
                   }_{\equiv \mt^{(mj)}(\bx,t)},
%     \specialnumber{a,b}
\end{align}
\end{subequations}
where both $\mw^{(mj)}(\bx,t)$ and $\mt^{(mj)}(\bx,t)$ are introduced by definition and they are real functions, with their nonzero complex envelopes $\bmw^{(mj)}$ given by, 
\begin{subequations}\label{eq:bmws}
\begin{align}
        \bmw\sz  = ~& \brwc\sz- \half (\nabla B_s+\rmi\bk_0B_s)^*\cdot(\nabla+\rmi\bk_0)A,  
        \\
		\bmw\ssd = ~& \brwc\ssd- \half(\nabla+\rmi\bk_0)B_s\cdot(\nabla+\rmi\bk_0)A,
		\\
		\bmw\trdf = ~& \brwc\trdf + \quard\p_zB^* \big[(\nabla+\rmi\bk_0)A\big]^2 +  \half \p_zB \big|(\nabla+\rmi\bk_0)A\big|^2,
		\\
		\bmw\trdt = ~& \brwc\trdt+\quard\p_zB \big[(\nabla+\rmi\bk_0)A\big]^2, 
		 \\
		\bmw^{(40)} = ~&\brwc^{(40)} + 
		                                     \half \brwc\sz \big|(\nabla+\rmi\bk_0)A\big|^2 
		                                     + \quard \brwc\ssd \big[(\nabla A+\rmi\bk_0 A)^*\big]^2 , 
		\\
		\bmw^{(42)} = ~&\brwc^{(42)} +  \half \mR\big[\brwc\sz \big](\nabla A+\rmi\bk_0 A)^2 +  \half \brwc\ssd \big|(\nabla+\rmi\bk_0)A\big|^2 , 
		\\
		\bmw^{(44)} = ~&\brwc^{(44)} + \quard \brwc\ssd (\nabla A+\rmi\bk_0 A)^2, 
\end{align}
\end{subequations}
for $z=0$, and the nonzero $\bmt^{(mj)}$ given by, 
\begin{subequations}\label{eq:bmts}
	\begin{align}
	\bmt\sz  = ~& -\quard |(\nabla+\rmi\bk_0)B|^2 +\quard |\p_zB|^2,
	\\
	\bmt\ssd = ~& -\quard\big[(\nabla+\rmi\bk_0)B\big]^2 +\quard (\p_zB)^2,
	\\
	\bmt^{(31)} = ~& \half \brwc\ssd\p_zB^* + \mR\big(W\sz \big)\p_zB , 
	\\
	\bmt\trdt = ~&  \half \brwc\ssd\p_zB , 
    \\
   \bmt^{(40)} = ~&\half \mR(\brwc\sz )^2+ \quard |\brwc\ssd|^2
               + \half \brwc\fst(\brwc\trdf)^* + \dfrac{1}{8}|\brwc\fst|^2|(\nabla+\rmi\bk_0)A|^2\notag\\
   ~& +  \dfrac{1}{16}(\brwc\fst)^2[(\nabla A+\rmi\bk_0 A)^*]^2  , 
   \\
   \bmt^{(42)} = ~& \mR\big[\brwc\sz \big] \brwc\ssd + \half \brwc\fst\brwc\trdf +\half (\brwc\fst)^*\brwc\trdt 
              \notag \\
              ~& + \dfrac{1}{8} |\brwc\fst|^2\big[(\nabla+\rmi\bk_0)A\big]^2 
                   + \dfrac{1}{8} (\brwc\fst)^2|\nabla A+\rmi\bk_0A|^2  , 
   	\\
   \bmt^{(44)} = ~&  \quard( \brwc\ssd )^2 + \half \brwc\fst\brwc\trdt +\dfrac{1}{16} (\brwc\fst)^2 (\nabla A+\rmi\bk_0A)^2, 
	\end{align}
\end{subequations}
\end{linenomath}
for $z=0$. Using envelope $A$ and $B_s$ as input, envelopes $B_0\mj$, $\brW\mj$, $\bmw\mj$, and $\bmt\mj$ are obtained in sequence from the lowest to higher orders in wave steepness, which will be directly used in the CEEEs derived in the following section. 

%---%---%----%----%----%
%            section         %
%---%---%----%----%----%
\subsection{The coupled envelope evolution equations (CEEEs)} 
\label{sub:ceees}
The $M-$th order accurate (in wave steepness) CEEEs are obtained through the following sequential procedures; (i)
inserting wave parameters and forcing terms in a form of the separation of wave harmonics presented in \S \ref{sub:sep_harm} into the evolution equations \specialeqref{eq:ode_zetapsi}{a,b}, (ii) keeping the components in the Fourier wavenumber region where $\bk\cdot\bk_0> 0$, and (iii) multiplying all terms by a factor of $\exp(-\rmi\alpha\bk_0\cdot\bx+\rmi\beta\omega_0t)$. Hence, the CEEEs are obtained
\begin{linenomath}
\begin{equation}\label{eq:ceees}
    ( \p_t-\rmi\beta\omega_0) A -  \p_zB   = ~ \mn_{A,M}
    ~\text{and}~
     ( \p_t-\rmi\beta\omega_0) B_s + g A=~ \mn_{B,M},
     \specialnumber{a,b}
\end{equation}
\end{linenomath}
with the terms for the complex conjugates removed, and the nonlinear forcing terms on the right side of the equations are given by
\begin{linenomath}
\begin{subequations}
\begin{align}
		\mn_{A,M}(\bx,t) =~ &  \sum\limits_{m=1}^{m=M} \mn_A^{(m) }
		 \equiv \sum\limits_{m=1}^{m=M} 
		\sum\limits_{j=0}^{j=m} \mn_{A}^{(mj)}
		\rme^{\rmi(j-1)(\alpha \bk_0\cdot\bx-\beta \omega_0t)},
		~\text{and}~            
		 \\             %
		\mn_{B,M}(\bx,t) = ~&   \sum\limits_{m=1}^{m=M} \mn_B^{m) }
		\equiv  \sum\limits_{m=1}^{m=M} 
		\sum\limits_{j=0}^{j=m}
		\mn_{B}^{(ij)}
		\rme^{\rmi(j-1)(\alpha \bk_0\cdot\bx-\beta \omega_0t)},
\end{align}
\end{subequations}
where $\mn^{(mj)}_A =0$ and $\mn^{(ij)}_B=0$ for $ij=10$ and $ij=11$, and the other non-vanishing components are expressed as
\begin{equation}
	 \mn_{A}^{(mj)} =
	 \big[ \mw^{(mj)}\big]^{[j]}_+	 
	 ~\text{and}~   
	 \mn_{B}^{(mj)} =
	 \big[ \mt^{(mj)}\big]^{[j]}_+,
	 \specialnumber{a,b}
\end{equation}
where, with $m=j$, the following relations hold
\begin{equation}
\mn_{A}^{(mm)} =
\bmw^{(mm)}
~\text{and}~   
\mn_{B}^{(mm)} =
\bmt^{(mm)},
\specialnumber{a,b}
\end{equation}
\end{linenomath}
attributing to that both $\bmw^{(mj)}$ and $\bmt^{(mj)}$ with $m=j$ are the product of the terms which are only nonzero in the half wavenumber plane where $(\bk+j\bk_0)\cdot\bk_0>0$ by definition, e.g., $\hat{B}\ssd(\bk,t)$ and $\hat{B}\trdt(\bk,t)$. We will see in \S\ref{sub:bound} that the waves forced by the nonlinear terms with $m=j$ can not be free. The exploration of the newly derived CEEEs given by \specialeqref{eq:ceees}{a,b}  in this paper are presented in \S\ref{sec:ceeesFeatures} and \S\ref{sec:comprisons}. 
%

%-------------%---%----%----%----%--------------%
%                    section                    %
%-------------%---%----%----%----%--------------%
%
\section{Discussion of the CEEEs}
\label{sec:ceeesFeatures}
The CEEEs given by \specialeqref{eq:ceees}{a,b} have a few key features which cannot be fully explored in this paper. In this section, three aspects are especially highlighted. The first is associated with their potential of  high numerical efficiency through using an exponential integrator, as examined in \S\ref{sub:expInt} and also \S\ref{sub:comp}.  Secondly, due to that the two main unknowns in the CEEEs are a pair of canonical variables, it is shown in \S\ref{sub:wae} that the CEEEs can lead to the nonlinear evolution equation of the wave action.  The third illustrates the clear physical meanings of the nonlinear forcing terms (i.e. $\mn_{A}^{(mj)}$ and $\mn_{B}^{(mj)}$ ) of different harmonics in  \S \ref{sec:nonLwaves} in their capability in the nonlinear forcing of different waves.  
\subsection{Analytical solution and numerical implementation using an exponential integrator}
\label{sub:expInt}
For the solution of the CEEEs, there are many applicable time integration methods. In this paper, we propose to using an exponential integrator, see, e.g., \cite{hochbruck10} for the details. This choice is made due to two aspects: (i) the terms involved in the CEEEs can be highly oscillatory and (ii) we can easily identify the terms with a highly oscillatory nature from these that are slowly varying. Preforming a Fourier transform on both sides of the CEEEs gives rise to
\begin{linenomath}
\begin{align} 
\label{eq:ceees_fft}
\left[
\begin{array}{c}
\dot{\hat{A}} \\
\dot{\hat{B}}_s
\end{array}
\right] 
= 
\left[
\begin{array}{cc}
\rmi\beta\omega_0  & |\bk+\bk_0|\tanh |\bk+\bk_0|h \\
- g  &   \rmi\beta\omega_0 
\end{array}
\right]
\left[
\begin{array}{c}
\hat{A} \\
\hat{B}_s
\end{array}
\right]
+ 
\left[
\begin{array}{c}
\hat{\mn }_{A,M}(\bk,\tau)\\
\hat{\mn }_{B,M}(\bk,\tau)
\end{array}
\right],
\specialnumber{a,b}
\end{align}
in which the dot denotes the derivative with respect to the time.  Following an exponential integrator, the analytical solution of \eqref{eq:ceees_fft} can be expressed in a form as
\begin{subequations}
\begin{align}\label{eq:ABs_sol}
\left[
\begin{array}{c}
{\hat{A}}(\bk,t) \\
{\hat{B}}_s(\bk,t)  
\end{array}
\right] 
= \bme((t-t_0)\Omega)
\left[
\begin{array}{c}
{\hat{A}}(\bk,t_0) \\
{\hat{B}}_s(\bk,t_0)  
\end{array}
\right]  + 
\sum\limits_{m=1}^{m=M} 
\sum\limits_{j=0}^{j=m} 
\left[
\begin{array}{c}
\mathcal{I}^{(mj)}_A(\bk,t)\\
\mathcal{I}^{(mj)}_B(\bk,t)
\end{array}
\right],
\end{align}
where the initial value problem is considered at the initial time instant $t_0$ when the envelopes are given, $ \bme$ denotes a matrix exponential given by
\begin{align}
\bme((t-t_0)\Omega) 
=~& 
\rme^{\rmi\beta\omega_0(t-t_0)} 
\left[
\begin{array}{cc}
\cos \big( (t-t_0)\Omega \big)   
& \dfrac{\Omega}{g} \sin \big((t-t_0)\Omega \big)
\\
- \dfrac{g}{ \Omega}  \sin \big((t-t_0)\Omega \big)
&\cos \big((t-t_0)\Omega\big)   
\end{array}
\right],
\end{align}
with $\Omega= \omega(\bk+\alpha\bk_0, h)$, and 
\begin{align}
\left[
\begin{array}{c}
\mathcal{I}^{(mj)}_A(\bk,t)\\
\mathcal{I}^{(mj)}_B(\bk,t)
\end{array}
\right]
=
\int\limits_{t_0}^t
   \bme((t-\tau)\Omega) 
    \rme^{-\rmi(j-1)\beta \omega_0\tau}
  \left[
  \begin{array}{c}
  \hat{\mn }^{(mj)}_{A}(\bk+(j-1)\alpha\bk_0,\tau)\\
  \hat{\mn }^{(mj)}_{B}(\bk+(j-1)\alpha\bk_0,\tau)
  \end{array}
  \right]
  \rmd \tau.
  \label{eq:Idefs}
\end{align}
\end{subequations}
 Similar to a NLS equation-based model, the analytical solution of the envelope evolution equations in a form as \eqref{eq:ABs_sol} consists of two terms: a linear and  nonlinear term corresponding to the first and the term of double sums on the right hand side of \eqref{eq:ABs_sol}, respectively.  \eqref{eq:ABs_sol} is obviously accurate for the evolution of linear surface waves. Owing to that the integrand of the integrals in \eqref{eq:Idefs} depends on the time-dependent envelopes, i.e., in an implicit form, the computation of \eqref{eq:ABs_sol}  requires a time integration method for the temporal-spatial evolution of nonlinear waves. 
 To this end, there are many available approaches, e.g., the midpoint  or the fourth-order Runge-kutta method.
 %  {\color{blue} The split-step approach for the numerical implementation of a NLS-based equation as used in \cite{lo85} stands for the same..... } The connections between the numerical aspects here are weak with the split-step approach due to the fact that a NLS equation is in a special form: $\p_tu=\rmi\omega_0\mL u +\alpha |u|^2 u$ .  
One would notice that 
a leading-order scale of  both $\hat{\mn}^{(mj)}_{A}$ and $\hat{\mn}^{(mj)}_{B}$ is $\mo(\eps^2)$ and their time derivative $\p_t\hat{\mn}^{(mj)}_{A}$ and $\p_t\hat{\mn}^{(mj)}_{B}$ have the scale $\mo(\eps^2\varepsilon_\tf)$, where $\varepsilon_\tf$ denotes the dimensionless frequency bandwidth. The aforementioned scale of analysis can be clearly demonstrated
in \S\ref{sub:nls_stokes}. Therefore, for updating \eqref{eq:Idefs} for one time step in the time interval $[t_n, t_n+\Delta t]$ with $t_n$ a  time instant when $\hat{A}(\bk,t_n)$ and $\hat{B}(\bk,t_n)$ were computed and $\Delta t$ a small time interval,  a numerical algorithm for time integration is required. The forward Euler method which is first order accurate in the temporal interval leads to
\begin{align}
\label{eq:Intn}
\left[
\begin{array}{c}
\mathcal{I}^{(mj)}_A(\bk,t)\\
\mathcal{I}^{(mj)}_B(\bk,t)
\end{array}
\right]
=
\mathbf{I}_n(t_{n+1}) 
\left[
\begin{array}{c}
\hat{\mn }^{(mj)}_{A}({\bk+(j-1)\alpha\bk_0},t_n)\\
\hat{\mn }^{(mj)}_{B}({\bk+(j-1)\alpha\bk_0},t_n)
\end{array}
\right]
%\int\limits_{t_n}^{t_{n+1}}
%\bme((t_{n+1}-\tau)\omega) 
%\rme^{-\rmi(j-1)\beta \omega_0\tau}
%\rmd \tau
+  \mo(\eps^2{\varepsilon_\tf}\varepsilon_t \omega_0\Delta t),
\end{align}
where $t_{n+1}=t_{n}+\Delta t$, $\varepsilon_t\sim 1/(\omega_0\Delta t)\ll 1$ such that the approximation to the time integration given by \eqref{eq:Intn} numerically converges, and the rapidly varying integral matrix $\mathbf{I}_n$ is defined as and thereby given by
\begin{subequations}
\begin{align}
      \mathbf{I}_n(t_{n+1}) \equiv & \int\limits_{t_n}^{t_{n+1}}
      \bme((t_{n+1}-\tau)\Omega) 
      \rme^{-\rmi(j-1)\beta \omega_0\tau}
      \rmd \tau 
      \\
    =  & -\half \rmi  \rme^{-\rmi(j-1)\beta \omega_0t_{n+1}}
      \left\{
       \dfrac{1-\rme^{\rmi [\omega_0+ (j-1)\beta\omega_0+\Omega]\Delta t}}{\omega_0+ (j-1)\beta\omega_0+\Omega } 
      \left[
      \begin{array}{cc}     1
      & \dfrac{\Omega}{g}
      \\
      -  \dfrac{g}
      { \Omega}  
      & 1
      \end{array}
      \right]
      + 
      \right.
      \notag\\
      %%%%%%%
      & ~~~~~~~~
      \left.
       \dfrac{1-\rme^{\rmi [\omega_0+ (j-1)\beta\omega_0-\Omega]\Delta t}}{\omega_0+ (j-1)\beta\omega_0-\Omega } 
      \left[
      \begin{array}{cc}     1
      & -\rmi \dfrac{\Omega}{g}
      \\
      \rmi \dfrac{g}
      { \Omega}  
      & 1
      \end{array}
      \right]
      \right\}. 
      \label{eq:In}
\end{align} 
\end{subequations}
\end{linenomath}
It is worth noting that a numerical algorithm more accurate than the forward Euler in the sense of time integration can be used to evaluate \eqref{eq:Idefs}. An exponential integrator can also be used for the HOS method following similar procedures in this section. 

Therefore,  through an exponential integrator and the forward Euler method, we obtain
\begin{linenomath}
\begin{align}\label{eq:hAB_tn}
\left[
\begin{array}{c}
{\hat{A}}(\bk,t_{n+1}) \\
{\hat{B}}_s(\bk,t_{n+1})  
\end{array}
\right] 
= & \bme((t_{n+1}-t_0)\Omega)
\left[
\begin{array}{c}
{\hat{A}}(\bk,t_0) \\
{\hat{B}}_s(\bk,t_0)  
\end{array}
\right]  + \\
&
\sum\limits_{m=1}^{m=M} 
\sum\limits_{j=0}^{j=m} 
\mathbf{I}_n(t_{n+1}) 
\left[
\begin{array}{c}
\hat{\mn }^{(mj)}_{A}(\bk+(j-1)\alpha\bk_0,t_n)\\
\hat{\mn }^{(mj)}_{B}(\bk+(j-1)\alpha\bk_0,t_n)
\end{array}
\right]\notag
 + \mo(\eps^2\varepsilon_t \omega_0\Delta t),
\end{align}
\end{linenomath}
which needs to be updated step by step from the initial  instant $t=t_0$ with given initial conditions. Compared with the HOS method implemented by \cite{ducrozet16} where the time interval depends on the shortest wave period, \eqref{eq:hAB_tn} permits a time interval which admits $\omega_0\Delta t\sim \mo(1/(\eps^2\varepsilon_\tf))$ for numerical stability and convergence and thereby a much larger value  without compromising the numerical efficiency with a careful choice of the value for $\beta$, as explained in \S\ref{sub:comp}. 
% {\color{red} (Different from the Zakharov equations, it does not involve zero denominators, etc. This should be somewhat mentioned.)}
%

%
%---%---%----%----%----%
%            section         %
%---%---%----%----%----%
%
\subsection{The energy balance equation}
\label{sub:wae}
Based on the CEEEs in the Fourier plane expressed as \eqref{eq:ceees_fft}, it is straightforward to derive  the energy balance equation. To this end,  the following parameter is introduced
\begin{align}\label{eq:hata_def}
    \hat{a} = \sqrt{\dfrac{g}{2\Omega}}\hat{A}
          + \rmi \sqrt{\dfrac{\Omega}{2g}} \hat{B},
\end{align}
which is a function of the same (slow) time as envelope $\hat{A}$ and $\hat{B}_s$. The following sequential procedures are used;  (i) Multiplying \specialeqref{eq:ceees_fft}{a} and \specialeqref{eq:ceees_fft}{b} by $\sqrt{g/(2\Omega)}$ and $\rmi \sqrt{\Omega/(2g)}$, respectively; (ii)  adding up the two resulting equations; (iii) replacing the terms which correspond to the definition of the new variable $\hat{a}$.  Thereby, we obtain
\begin{linenomath}
\begin{align}\label{eq:wae_mid}
    \p_t\hat{a} + \rmi(\Omega-\beta\omega_0) \hat{a} =
    \sqrt{\dfrac{g}{2\Omega}} \hat{\mn}_A
    + \sqrt{\dfrac{\Omega}{2g}}\hat{\mn}_B.
\end{align}
Multiplying \eqref{eq:wae_mid} and its complex conjugates by $\hat{a}^*$ and $\hat{a}$, respectively, and adding up the resulting equations leads to
\begin{subequations}
\begin{align}\label{eq:wae}
   \p_{t} ( \hat{a}\hat{a}^* )= \dfrac{g}{2\Omega} (\hat{\mn}_{B,M}^*\hat{a} + \hat{\mn}_{B,M}\hat{a}^*) 
   + 
   \dfrac{\Omega}{2g} (\hat{\mn}_{B,M}^*\hat{a} + \hat{\mn}_{B,M}\hat{a}^*),
\end{align}
in which $\hat{a}\hat{a}^*$ denotes the wave action similar to that defined in an extensive body of literature, e.g., \cite{zakharov68,stiassnie84b,krasitskii94,annenkov09,gramstad14}. Equation \eqref{eq:wae} is known as the energy balance equation or the wave action equation. With the nonlinear effects neglected, the  conservation of linear wave actions is evident
\begin{align}\label{eq:wae_conserv}
 \p_{t} ( \hat{a}\hat{a}^* )= 0,
\end{align}
\end{subequations}
\end{linenomath}
which  suggests that the transfer of wave actions does not occur between linear waves, as it should be. 
%

%---%---%----%----%----%
%        section       %
%---%---%----%----%----%
%
\subsection{Nonlinear forcing of waves}\label{sec:nonLwaves}
%
% {\color{blue} (The nonlinear terms have very clear physical meanings, which should be mentioned at least somewhere)}
%
Due to the  nonlinear effects which lead to the  forcing terms on the right hand side of the CEEEs described by \specialeqref{eq:ceees}{a,b}, it is understood that both  free and bound (locked)  waves, which do and do not obey the dispersion relation, respectively, can be forced.  The former can arise from  resonant and instability conditions which are the result of a combination of bandwidth and nonlinearity,  as studied by numerous works, noticeably \cite{phillips60,hasselmann62,benjamin67,zakharov68,longuet78}, and \cite{mclean82b,mclean82}. We show in this section how the nonlinear terms of different wave harmonics on the right hand side CEEEs lead to the forcing of bound waves and resonant free waves.

%Eliminating $B_s$ in \specialeqref{eq:ceees_stokes}{a,b} leads to the following equation 
%%
%\begin{align}\label{eq:nls_stokes}
%  \p_t A =\dfrac{(\p_t-\rmi\omega_0)}{ -2\rmi\omega_0} \mn_{A,3}+\dfrac{\rmi\omega_0}{2g}\mn_{B,3} ,
%\end{align}
%%
%{\color{blue} (double check this)} which should lead to the classic NLS equation \cite{zakharov68} obtained the NLS equation. 
%{\color{blue} (It should be mentioned that we can also recover to the NLS equation derived by \cite{li21nls}.)}

For later reference, we introduce the dimensionless wave vector $\be_n=\bk_n/k_0$ and wave frequency $\sigma_n=\omega(|\bk_n|h)/\sqrt{gk_0}$, which are given by, respectively
\begin{linenomath}
\begin{equation}
   \be_{n} = (1+p_n, q_n)~
   \text{and}~
   \sigma_n = \sqrt{\tanh (k_0|\be_n|h)}
   ~
   \text{for}~n\in\{1,2,3,...\},
   \specialnumber{a,b}
\end{equation}
where $n$ denotes the $n-$th free wave and $\be_n\cdot\be_0>0$ with $\be_0 =(1,0)$ and $p_n>-1$ and $q_n$ an arbitrarily chosen parameter. 
For an infinitesimal wave steepness, the resulting dimensionless wave vector, $\be_N$, and angular frequency, $\sigma_N$, due to  nonlinear forcing term $\mn_{A}\mj\rme^{\rmi(j-1)(\alpha \bk_0\cdot\bx-\beta \omega_0t)}$  and $\mn_{B}\mj\rme^{\rmi(j-1)(\alpha \bk_0\cdot\bx-\beta \omega_0t)}$, can be obtained through the analysis in the Fourier plane and the  superposition of linear waves. 
Especially, they are obtained by inserting the linear approximations of the unknown envelopes
\begin{subequations}
\begin{align}
    \hat{A}(\bk,t)\approx~& \hat{A}(\bk,t_0)\exp[\rmi\bk\cdot\bx-\rmi(\Omega-\beta\omega_0)(t-t_0)]
    ~\text{and}~
    \\
    \hat{B}_s(\bk,t)\approx~&
    \hat{B}_s(\bk,t_0)\exp[\rmi\bk\cdot\bx-\rmi(\Omega-\beta\omega_0)(t-t_0)],
\end{align}
\end{subequations}
into the envelopes of the potential and vertical velocity and thereafter the nonlinear forcing term of the $j$-th harmonic at $\mo(\eps^m)$; explicitly, we arrive at
\begin{subequations}\label{eq:ww_interac}
\begin{align}
    \be_{N} -\alpha \be_0=~& (j-1)\alpha\be_0 + \sum_{n=1}^m\pm( \be_{n}- \alpha \be_0 ),\\
    \sigma_{N} -\beta \sigma_0=~& (j-1)\beta\sigma_0 + \sum_{n=1}^m \pm( \sigma_n-\beta \sigma_0 ),
\end{align}
\end{subequations}
\end{linenomath}
where $\sigma_0=\sqrt{\tanh k_0h}$, $\be_N\cdot\be_0>0$ holds for nonvanishing $\hat{A}$ and $\hat{B}_s$ by definition, and  $N=m+1$  denotes the number of waves involved in the interaction. Which sign to choose between `$\pm$' depends on $j$; for $j=m$, the `+' sign needs to be taken for all $n$ values whereas it is not permitted to choose `$-$' sign for all $n$ values as it will lead to the inequality $\be_N\cdot\be_0<0$ where $\hat{A}$ and $\hat{B}_s$ vanish. We highlight that $N=3$, $N=4$, and $N=5$ correspond to triad, quartet, and quintet wave interactions, which occur at the second, third, and fourth order in wave steepness, respectively.  With $\alpha=0$ and $\beta=0$, one would readily see that \specialeqref{eq:ww_interac}{a,b} are simply the kernels arising from $N$-wave interaction based on the Zakharov integral equation \citep{stiassnie84b,shrira96,janssen09}, as it should be. 

Physically, the resonant condition for $N$-waves interaction means that the resulting wave vector and frequency obey the dimensionless linear dispersion relation
\begin{align}
    \sigma_N - \sqrt{\tanh (k_0|\be_N|h)} = 0,
\end{align}
which can be used in the analysis of the nonlinear forcing of free waves arising from the interaction between linear waves. Mathematically, it corresponds to the particular terms (i.e. the nonlinear forcing terms) of the non-homogeneous CEEEs have components which coincide to the eigenvalues of the homogeneous CEEEs in frequency, leading to a linear growth in time similar to the discussion by \cite{hasselmann62}. This point in principle determines whether the nonlinear forcing of waves are free or bound, which are discussed in \S\ref{sub:resonance} and \S\ref{sub:bound}m respectively. 

%---%---%----%----%----%
%       section        %
%---%---%----%----%----%
%
\subsubsection{Forcing of  free waves due to resonant effects}
\label{sub:resonance}
The nonlinear forcing of free waves due to the class I resonant condition occurs at third order in wave steepness
for waves of first harmonic and corresponds to the effects of the nonlinear forcing terms with $m=3$ and $j=1$ in the CEEEs.
Following the previous works, e.g., \citet{stiassnie84b}, the class I resonant condition  due to quartet (linear) wave interaction in the CEEEs is given by, 
\begin{linenomath}
\begin{equation}\label{eq:res_quart}
    \be_4 = \be_1+\be_2-\be_3~\text{and}~\sigma_4 = \sigma_1+\sigma_2-\sigma_3,
    \specialnumber{a,b}
\end{equation}
where  $    \be_1=(1+p, q) , ~\be_2= (|1-p|,-q), ~\be_3=\be_4\equiv (1,1)$, and $\sigma_4=\sigma_0$.  Similarly, the class II resonant condition due to quintet wave interaction occurs due to the nonlinear forcing terms with $m=4$ and $j=2$ in the CEEEs where
\begin{equation}\label{eq:res_quin}
\be_5 = \be_1+\be_2+\be_3-\be_4~\text{and}~\sigma_5 = \sigma_1+\sigma_2+\sigma_3-\sigma_4,
\specialnumber{a,b}
\end{equation}
\end{linenomath}
with $\be_5=(1,0)$, $\sigma_5=\sigma_0$. The  effect of wave nonlinerity on the dispersion relation are neglected in \specialeqref{eq:res_quin}{a,b}. In order to account for this, the `Stokes-corrected' nonlinear dispersion relation, which can be obtained from (2.21c) by \cite{stiassnie84b},  should be used for $\sigma_n$ with $n\in \{1,2,3,4,5\}$.

\subsubsection{Bound waves}
\label{sub:bound}
With $j=m$ for $m\geq 2$, the inequality $\be_n\cdot\be_0>0$ holds for all wave vectors in \specialeqref{eq:ww_interac}{a,b} due to the definition of the envelope transform and the expression of $B^{(nn)}$ and $A$. This can be demonstrated by a simple example. To this end, we  choice the third term in \specialeqref{eq:wdefs}{g}.  % as $\brW\trdt$  appears directly in the expression \eqref{eq:bmws} for $\bmw\trdt$ given that $\bmw\trdt=\mn_{A}\trdt$. 
 It is understood that
\begin{linenomath}
\begin{align}\label{eq:nm33_1}
     \dfrac{1}{8}A^2\p_{zzz}B 
     \rme^{2\rmi(\alpha\bk_0\cdot\bx-\beta\omega_0t)}
     \equiv &
     % \dl{\dfrac{1}{64\pi^3}}
     \int
     \hat{A}(\bk_1,t)\hat{A}(\bk_2,t)|\bk_3+\bk_0|^3 \tanh(|\bk_3+\bk_0|h) 
     \times
     \notag
     \\
   & \hat{B}_s(\bk_3,t) \rme^{\rmi(\bk_1+\bk_2+\bk_3+2\alpha\bk_0-2\beta\omega_0t)\cdot\bx}\rmd \bk_1\rmd\bk_2\rmd\bk_3,
\end{align}
where $\bk_n+\bk_0 =k_0\be_n$ and $(\bk_n+\bk_0)\cdot\bk_0>0$ for non-vanishing $\hat{A}(\bk_n,t)$ and $\hat{B}_s(\bk_n,t)$ by  definition for $n=1,~2$, and $n=3$. The frequency superposition can be obtained through  the linear approximation to the CEEEs:
\begin{align}\label{eq:hAB_L}
		[\hat{A}(\bk_n,t) , \hat{B}_s(\bk_n,t) ]=[\hat{A}(\bk_n,t_0) , \hat{B}_s(\bk_n,t_0) ]\rme^{-\rmi(\omega_n-\beta\omega_0 )(t-t_0)} + \mo(\eps^2). 
\end{align}
where 
$\omega_n= \omega(\bk_n+\alpha\bk_0,h)$.
Inserting \eqref{eq:hAB_L} into \eqref{eq:nm33_1} and taking the frequency superposition readily leads to the frequency combination on the right hand side of \specialeqref{eq:ww_interac}{b}. The last step is to multiply the factor $\exp[\rmi(\alpha\bk_0\cdot\bx-\beta\omega_0t)]$ on both sides of the resulting equation and hence, we arrive at
\begin{align}\label{eq:nm33_2}
\dfrac{1}{8}A^2\p_{zzz}B 
\efactrd
\equiv &
%\dl{\dfrac{1}{64\pi^3}}
\int
\hat{A}(\bkp_1-\alpha\bk_0,t_0)\hat{A}(\bkp_2-\alpha\bk_0,t_0)\hat{B}_s(\bkp_3-\alpha\bk_0,t_0)
\times
\notag
\\
|\bkp_3|^3 \tanh(|\bkp_3|h)  & \rme^{%\left[
	\rmi \sum\limits_{n=1}^3\left[\bkp_n\cdot\bx-\sqrt{g|\bkp_n|\tanh(|\bkp_n|h)}(t-t_0) \right]%  \right]
}\rmd \bkp_1\rmd\bkp_2\rmd\bkp_3,
\end{align}
where the relation $\bkp_n=\bk_n+\alpha\bk_0$ was used for the change of the integral variables.  It becomes evident that the nonvanishing integrand in \eqref{eq:nm33_2} leads to the resulting waves whose dimensionless wave vector and frequency are given by
\begin{align}
  \be_4 = \sum\limits_{n=1}^{3}\bkp_n/k_0~\text{and}~\sigma_4 = \sum\limits_{n=1}^{3}  \sqrt{g|\bkp_n|\tanh(|\bkp_n|h)} /\omega_0
  ~\text{with}~
  \bkp_n\cdot\bk_0>0,
\end{align}
\end{linenomath}
which can be written in a form similar to \specialeqref{eq:res_quart}{a,b}, except that the third wave on the right hand side needs to take the positive sign instead. Due to this, the resonant quartet wave condition cannot be satisfied. Thereby, the nonlinear term chosen as an example can only lead to the forcing of bound waves which do not obey the linear dispersion relation. Similar analysis can be carried out for the other remained components of the nonlinear forcing terms with $m=j$. The third and fourth wave given by the quintet and quartet resonant condition in \eqref{eq:res_quart} and \eqref{eq:res_quin}, respectively, needs to take a negative sign. In contrast, all waves in the nonlinear terms on the right hand of the CEEEs with $m=j$ can only take the positive sign as the inequality $\be_n\cdot\be_0>0$ should hold as noted. Therefore, it can be readily inferred that only bound waves can be forced by the nonlinear terms in the CEEEs with $m=j$.

The discussions above in this section have covered the forcing of either free or bound waves arising from the nonlinear forcing terms in the CEEEs but these for $j=0$ and $m\geq 2$ which only appear in the even-th orders in wave steepness and which are often  responsible for the forcing of mean flows (or waves with a  low or vanishing frequency). These nonlinear forcing terms $j=0$ cannot force free waves since their resulting wave vectors and wave frequencies would not obey the resonant conditions presented in \S\ref{sub:resonance} for the interaction of infinitesimal waves. They are also associated with the singular terms in the Zakharov's kernel functions, as clearly stated in the introduction of \citet{gramstad14}.  Evidently,  nonlinearity effects on the dispersion relation have been neglected  in this section, and thereby how they affect the forcing of mean flows has not  been  explored.  At the second order with `$mj=20$',  we understand that subharmonic bound waves can be forced, as is well known, see, e.g., \cite{phillips60,hasselmann62}. The author conjectures that novel physics may be elucidated through the exploration of these  terms (which can force nonlinear mean flows) with the consideration of higher-order nonlinearity. As it is not the main focus of this paper, this aspect will be left for future explorations.

%
%-------------%---%----%----%----%--------------%
%                    section                    %
%-------------%---%----%----%----%--------------%
%
\section{Comparisons with two other methods}\label{sec:comprisons}
For further demonstrating the potential, the CEEEs are firstly compared with  a traditional perturbation expansion for the evolution of a train of Stokes waves (\S\ref{sub:nls_stokes}) and irregular waves with an arbitrary bandwidth and directional spreading (\S\ref{sub:ceees_li21}). Next, we proceed to comparisons with the HOS method in \S\ref{sub:hoscomp} of the nonlinear forcing terms in a limiting case (\S\ref{sub:num_nonlinear}) and of the computational complexity (\S\ref{sub:comp}) to especially demonstrate the numerical efficiency of the CEEEs. 
%

%---%---%----%----%----%
%       section        %
%---%---%----%----%----%
%
\subsection{Relation with a traditional perturbation expansion} 
\label{sub:comp_pertb}
Using an example of both a train of Stokes waves in \S\ref{sub:nls_stokes} and the more general weakly nonlinear three-dimensional waves in \S\ref{sub:ceees_li21}, we show in this section how to establish the relation between a traditional perturbation method and the CEEEs. The analytical analysis in the section has a twofold sub-goal. It firstly demonstrates that the CEEEs are correctly derived. Secondly, it shows the first few steps which are essential to more general derivations for bridging the relations between the CEEEs and other higher-order frameworks, e.g., the different versions of third-order NLS equations. For example, if both the third orders in $\eps_0$ and a narrow bandwidth are additionally considered in \S\ref{sub:nls_stokes}, the classic third-order accurate NLS equation would be recovered based on the CEEEs. Or if the Stokes waves are considered up to the fifth order in $\eps_0$ in  \S\ref{sub:nls_stokes}, one would deduce the framework by \cite{fenton85} starting from the CEEEs.
\subsubsection{A train of Stokes waves}
\label{sub:nls_stokes}
%
% For the detailed derivations of this section, refer to the nete_eqs.pdf
%
% We proceed to show how to establish the relationship between a classic perturbation expansion and the CEEEs for the evolution of Stokes waves. 
A train of Stokes waves is considered to have a wave vector of $\bk_0$ and  phase of $\theta_0$ and we choose $\alpha=1$ and $\beta=1$ for the implementation of the CEEEs. In a traditional perturbation method as noted in \S\ref{sub:tra_pertb}, it is typical of solving for the wave-perturbed parameters on a still water surface, in contrast to the CEEEs method where the primary unknowns are these defined on the free water surface. In the CEEEs, envelope $A=A(\bx,t)$ and $B_s=B_s(\bx,t)$ depend on both time and the horizontal position vector. They can be expressed in a form of power series of the wave steepness $\eps_0$, up to the second order
\begin{linenomath}
\begin{equation}\label{eq:AB_stokes}
    A =  \eps_0 A_1 + \eps_0^2 A_2\efacff 
    ~\text{and}~
    B_s = \eps_0 B_{s,1} + \eps_0^2 B_{s,2}\efacff,
    \specialnumber{a,b}
\end{equation}
%
% \begin{subequations}\label{eq:AB_stokes}
% \begin{align}
%     A = ~& \eps_0 (A_{11}+A_{31}) + \eps_0^2 (A_{22} + A_{42} )\efacff +\eps_3 \efactf
%     \\
%     B_s =~& \eps_0 B_{1} + \eps_0^2 B_{s,2}\efacff,
% \end{align}
% \end{subequations}
%
where the constant mean of both the envelope and the potential are neglected for simplicity but can be additionally considered;  $\eps_0$ denotes the small dimensionless wave steepness in a traditional perturbation method as noted in \S\ref{sub:tra_pertb}; subscript `1' and `2' denotes the envelopes at first and second-order in wave steepness, $\eps_0$, respectively;  $A_1=A_1(\eps_0^2t)$ and $B_{s,1}=B_{s,1}(\eps_0^2t)$ are a real and imaginary time-dependent amplitude of the linear elevation and the potential on a still water surface in a traditional perturbation method, respectively (see, e.g., \cite{fenton85}).   Similarly, a leading-order approximation to envelope $\brwc\fst$ is, to $\mo(\eps^2_0)$
\begin{equation}\label{eq:brwc_stokes}
    \brwc\fst = k_0\left(\eps_0\tanh (k_0h) B_{s,1} + 2\eps_0^2\tanh 2k_0h B_{s,2}\efacff  \right),
\end{equation}
where the factor of $2$ arises from the derivative with respect to the vertical axis due to the second-order superharmonic waves. 
Inserting \specialeqref{eq:AB_stokes}{a,b} and \eqref{eq:brwc_stokes} into the elevation, the potential on the free water surface, and the velocity potential at an arbitrary depth in the framework of the CEEEs leads to, up to second order in wave steepness $\eps_0$,
\begin{subequations}\label{eqs:ceeeStokes}
\begin{align}
    \zeta(\bx,t) = ~& \half\eps_0 A_1\efacff +\half \eps_0^2 A_2\efactf+ \cc
    % ~\text{and}~
    \\
    \psi(\bx,t) = ~& \half \eps_0 B_{s,1}\efacff + \half \eps_0^2 B_{s,2}\efactf +\cc
    \\
    \Phi(\bx,z,t) = ~& \half \eps_0 B_{s,1}
    \dfrac{\cosh k_0(z+h)}{\cosh k_0h}\efacff 
    \\
     &  +\half \eps_0^2\left(B_{s,2}- \half
     k_0\tanh k_0h A_1 B_{s,1}\right)\dfrac{\cosh 2k_0(z+h)}{\cosh 2k_0h} \efactf,
     \notag
\end{align}
\end{subequations}
Therefore, the CEEEs for the evolution of a train of Stokes waves can be much simplified to 
\begin{equation}\label{eq:ceees_stokes}
 (\p_t-\rmi\omega_0) A - \brwc\fst = \mn_{A,2}
 ~\text{and}~
  (\p_t-\rmi\omega_0) B_s + g A = \mn_{B,2}, 
  \specialnumber{a,b}
\end{equation}
where
\begin{equation}
        \mn_{A,2} = \mn_A\ssd \efacff
        ~\text{and}~
       \mn_{B,2} = \mn_B\ssd \efacff,
       \specialnumber{a,b}
\end{equation}
and
\begin{subequations}\label{eqs:mnAB_22stokes}
\begin{align}
     \mn_A\ssd = ~& \half \left(
    1- 2\tanh k_0h \tanh 2k_0h
   \right)k_0^2 A_1B_{s,1}
   \\
    \mn_B\ssd  = ~&\quard k_0^2(1+\tanh^2k_0h ) B_{s,1}^2.
\end{align}
\end{subequations}
Inserting $A_1=A_1(\eps_0^2t)$ and $B_{s,1}=B_{s,1}(\eps_0^2t)$ into \specialeqref{eqs:mnAB_22stokes}{a,b} and performing the analysis of scales gives rise to
\begin{subequations}
\begin{align}
   \mo\left(
   \mn\ssd_A, \mn\ssd_B 
   \right) \sim \mo(A_1B_{s,1}) \sim \eps_0^2
   ~\text{and}~ \\
   \mo\left(
   \p_t\mn\ssd_A, \p_t\mn\ssd_B
   \right)
   \sim \mo(\eps_0\p_t A_1)
   \sim\mo(\eps_0\p_t B_{s,1})
   \sim  \eps_0^4,
\end{align}
\end{subequations}
and therefore, $\mn_{A}^{(mj)} = \mn_{A}^{(mj)}(\eps_0^2t) $ and $\mn_{B}^{(mj)}=\mn_{B}^{(mj)}(\eps_0^2t)$, which suggests $\varepsilon_\tf=\eps_0^2$ as is well known for the  temporal evolution of the amplitude of a train of Stokes waves \citep{zakharov68,fenton85}, e.g., $A_1\equiv A_1(\eps^2_0t)=A_1(\varepsilon_\tf t) $ as noted. A traditional perturbation method proposes to solve the equations in sequence from the first to higher orders in wave steepness $\eps_0$ and thereby, we obtain
\begin{equation}
  \mo(\eps_0): -\rmi\omega_0 A_1 - k_0\tanh k_0h B_{s,1} = 0
  ~\text{and}~ -\rmi\omega_0 B_{s,1} - gA_1 = 0,
  \specialnumber{a,b}
\end{equation}
which are well known for the linear evolution of  monochromatic waves; at $ \mo(\eps_0^2)$
\begin{subequations}\label{bvps:eps0sq}
\begin{align}
   A_2\efacff  -\dfrac{\rmi k_0\tanh 2k_0h}{\omega_0} B_{s,2}\efacff = ~& \dfrac{\rmi}{2\omega_0} \mn_A\ssd\efacff,
     \\
   2\rmi\omega_0 B_{s,2}\efacff + gA_2\efacff = ~& - \mn\ssd_B\efacff,
\end{align}
\end{subequations}
where, despite that it can be eliminated, the factor $\efacff$ was kept with the intention to demonstrate clearly that the numerical solution of  \specialeqref{bvps:eps0sq}{a,b} for the unknown parts of envelope $A$ and $B_s$, i.e., $A_2\efacff$ and $B_{s,2}\efacff$, depends only on the temporal scale of the forcing terms $\mn\ssd_A$ and $\mn\ssd_B$. Especially in the limiting case of second-order Stokes waves, the solution of \specialeqref{bvps:eps0sq}{a,b} obtained numerically for bears no mathematical truncation error \citep[\S 1.3]{atkinson08} as the unknown envelopes are approximately constants, and therefore the numerical solution can return the same results as the analytical solution from trivial algebras. Comparing the expressions given by \specialeqref{eqs:ceeeStokes}{a--c} in the CEEEs and these by \specialeqref{eq:sfbc_li21}{a,b} for the same parameter, the following relations hold 
\begin{equation}\label{stokes:AB_ABprime}
    A'_{mm} = A_{m}, ~B'_{11} = B_{s,1}, 
    ~\text{and}~
    B'_{22} = B_{s,2}  -\half k_0\tanh k_0h A_{1}B_{s,1},
    \specialnumber{a,b,c}
\end{equation}
\end{linenomath}
where $m=1$ and $m=2$. Replacing the parameters in \specialeqref{bvps:eps0sq}{a,b} with these used in a traditional perturbation method through the relations \specialeqref{stokes:AB_ABprime}{a,b,c},  eliminating the factor $\efacff$,  and using the relation $(\p_{t}-\rmi\omega_0)A_{11}' =\p_{z}B'_{11}$ readily leads to  the envelope equations given by \specialeqref{eq:ABp_22}{a,b}. Hence, the relation between the CEEEs and a traditional perturbation method for the evolution of a train of weakly nonlinear Stokes waves has been established.

% one would recover the 
% similar to . presented here can be easily extended to the third and higher orders for the nonlinear evolution of Stokes waves. Moreover, one would recover the NLS equation for both the evolution. 
% %

%---%---%----%----%----%
%             section             %
%---%---%----%----%----%
%
\subsubsection{The semi-analytical framework for directionally-spread broadband waves}
\label{sub:ceees_li21}
Similar to \S\ref{sub:nls_stokes}, the main focus of this section is to show how the CEEEs leads to the framework by \cite{lili21} for the evolution of three-dimensional broad-band waves with large directional spreading. Again, we start from  envelope $A$ and $B_s$  in the CEEEs in a form of power series in $\eps_0$ up to the second order in $\eps_0$
\begin{linenomath}
\begin{subequations}\label{eq:AB_eps0}
\begin{align}
    A = ~& \eps_0 A_{11} +\eps^2_0 A_{20}\rme^{-\rmi(\alpha\bk_0\cdot\bx-\beta\omega_0t)}+ \eps_0^2 A_{22}\efac,\\
    B_s = ~& \eps_0 B_{s,11} + \eps^2_0 B_{s,20}\rme^{-\rmi(\alpha\bk_0\cdot\bx-\beta\omega_0t)}+ \eps_0^2 B_{s,22}\efac,
    % , B\ssd = -\half \p_z B_1 A_s
\end{align}
\end{subequations}
\end{linenomath}
where $A_{mj}=A_{mj}(\bx,t)$ and $B_{s,mj}=B_{s,mj}(\bx,t)$. The relationship between \specialeqref{eq:AB_eps0}{b} and the envelope of the potential and vertical velocity on the free water surface leads to
\begin{linenomath}
\begin{subequations}\label{eq:BWs_eps0}
\begin{align}
    B(\bx,z,t) =~& \eps_0 B_{11}(\bx,z,t)+ \eps_0^2 B_{20} \rme^{-\rmi(\alpha\bk_0\cdot\bx-\beta\omega_0t)} 
       + \eps_0^2  B_{22}\efac
       \\
   \brwc\fst(\bx,t) = ~& \eps_0 \brwc\fst_{11}(\bx,0,t)+ 
   \eps_0^2 \brwc\fst_{20}(\bx,0,t)
   \rme^{-\rmi(\alpha\bk_0\cdot\bx-\beta\omega_0t)} \notag\\
      &  +\eps_0^2  \brwc\fst_{22}(\bx,0,t)\efac,
\end{align}
with $B_{s,mj}(\bx,t) = B_{mj}(\bx,0,t)$ and
\begin{align}
    B_{mj} =~&  \kint \hat{B}_{s,mj}
    \dfrac{\cosh |\bk+j\alpha \bk_0|(z+h)}{\cosh |\bk+j\alpha \bk_0|h}\rme^{\rmi\kx}\rmd\bk,
    \\
    \brwc\fst_{mj}(\bx,z,t) =~& \p_zB_{mj} \equiv  \kint |\bk+j\alpha \bk_0|\hat{B}_{s,mj}\dfrac{\sinh |\bk+j\alpha \bk_0|(z+h)}{\cosh |\bk+j\alpha \bk_0|h}\rme^{\rmi\kx}\rmd\bk.
\end{align}
\end{subequations}
We next derive the parameters needed at second order in wave steepness in the CEEEs. Substituting the envelopes on the still water surface given by  \specialeqref{eq:AB_eps0}{a,b} and \specialeqref{eq:BWs_eps0}{a,b} into the envelopes of the approximate wave fields at the second order in wave steepness presented in \S\ref{sub:cees_orderM}, and keep the terms up to $\mo(\eps_0^2)$ gives rise to
\begin{subequations}
\begin{align}
    B_0\sz  =~& - \eps_0^2\left[\mR\left(\half A^*_{11}B_{s,11}\right) \right]^{[0]}_+,
    \\
    B_0\ssd = ~& -\half \eps^2_0 A_{11}B_{s,11}, 
    \\
    \brwc\ssd = ~&  \eps_0^2\kint |2(\bk_0+\bk)| \hat{B}_0\ssd \dfrac{\sinh |(2\alpha\bk_0+\bk)(z+h)|}{\cosh |2(\bk_0+\bk)h|}\rme^{\rmi\kx}\rmd\bk
    + \half  \eps_0^2 A_{11} \p_z\brwc\fst_{11},
    \\
     \brwc\sz = ~& \eps_0^2 \kint |\bk| \hat{B}_0\sz \dfrac{\sinh |\bk|(z+h)}{\cosh |\bk| h}\rme^{\rmi\kx}\rmd\bk
    + \eps_0^2\mR \left(\half A^*_{11} \p_z\brwc\fst_{11} \right).
\end{align}
\end{subequations}
Similarly, inserting the resulting expressions for the envelopes in the first and second orders in this section into the nonlinear forcing terms in the second order in the CEEEs  leads to
\begin{subequations}
\begin{align}
    \mn_A\ssd = ~& 
     \eps_0^2 \kint 
     |2(\bk_0+\bk)| \hat{B}_0\ssd \dfrac{\sinh |(2\alpha\bk_0+\bk)(z+h)|}{\cosh |2(\bk_0+\bk)h|}\rme^{\rmi\kx}\rmd\bk
    \notag \\ 
    +&\eps_0^2 \left[\half  A_{11} \p_z\brwc\fst_{11} -\half (\nabla+\rmi\bk_0)A_{11}\cdot(\nabla+\rmi\bk_0)B_{s,11}    \right],
    \\
    \mn_A\sz = ~& 
    \eps_0^2 \kint |\bk| \hat{B}_0\sz \dfrac{\sinh |\bk|(z+h)}{\cosh |\bk| h}\rme^{\rmi\kx}\rmd\bk,
    \\
    +& \eps_0^2 \left[
    \mR \left(\half A^*_{11} \p_z\brwc\fst_{11}
    -\half (\nabla-\rmi\bk_0)A^*_{11}\cdot(\nabla+\rmi\bk_0)B_{s,11} \right)\right]_+^{[0]},
    \notag
    \\
    \mn_B\ssd = ~& \quard \left[-(\nabla B_{s,1}+\rmi\bk_0B_{s,1})^2 +(\brwc_{11}\fst)^2 \right],
    \\
    \mn_B\sz = ~& \quard \left[-|\nabla B_{s,1}+\rmi\bk_0B_{s,1}|^2 +|\brwc_{11}\fst|^2 \right]_+^{[0]}. 
\end{align}
\end{subequations}
\end{linenomath}
Inserting the parameters in a form of power series of $\eps_0$ back to the CEEEs,  collecting the terms in the second order in $\eps_0$, and separating the wave harmonics leads to the evolution equations for the second-order super-harmonic waves
\begin{linenomath}
\begin{equation}\label{eq:AB_22}
    (\p_t-2\rmi\beta\omega_0) A_{22} - \brwc\fst_{22} = 
    \mn\ssd_A
    ~\text{and}~
    (\p_t-2\rmi\beta\omega_0) B_{s, 22} + g A_{22}  = 
    \mn\ssd_B ,
    \specialnumber{a,b}
\end{equation}
\end{linenomath}
with the factor $\efac$ eliminated on both sides of the equation,
and the equations for the subharmonic waves
\begin{equation}\label{eq:AB_20}
    \p_t A_{20} - \brwc\fst_{20} = 
    \mn\sz_A
    ~\text{and}~
   \p_t B_{s, 20} + g A_{20}  = 
    \mn\sz_B. 
     \specialnumber{a,b}
\end{equation}
with the factor $\exp{[-\rmi(\alpha\bk_0\cdot\bx-\beta\omega_0t)]}$ eliminated. The next step is to establish the relationship between $B_{s,22}$ and $\brwc\fst_{22}$ and the corresponding parameter given by \cite{lili21}, respectively. Comparing  \specialeqref{eq:AB_eps0}{a} for the envelope of the elevation in the CEEEs and that given by \specialeqref{eq:sfbc_li21}{c}  in the framework by Li \& Li (2021), it is evidently understood that $A_{mj}= A_{mj}'$ and
\begin{subequations}\label{eqs:li21}
\begin{align}
    B'_{11}(\bx,0,t)=~& B_{11}(\bx,0,t)=B_{s,11}, 
    \\
    B'_{22}(\bx,0,t) =~&  B_{s,22} - \half A'_{11}\p_zB'_{11},
    \\
    B'_{20}(\bx,0,t) = ~& B_{s,20}  -\left[\mR\left(\half A^{'*}_{11}B'_{11}\right) \right]^{[0]}_+, 
    \\
    \p_zB'_{22}(\bx,0,t) = ~& \brwc\fst_{22} + \p_zB\ssd(\bx,0,t),
    \\
    \p_zB'_{20}(\bx,0,t) = ~& \brwc\fst_{20} +\p_zB\sz(\bx,0,t).
\end{align}
\end{subequations}
Replacing the terms in the CEEEs with these by \cite{lili21}  through the connections given by \specialeqref{eqs:li21}{a--e} leads to the envelope equations of \cite{lili21} given by \specialeqref{eq:sfbc_li21}{a,b}, for which the identity $(\p_t-\rmi\omega_0)A'_{11}= \p_zB'_{11}  $ from the linearized kinematic condition on a still water surface was used.

Based on \specialeqref{eqs:li21}{a--e} and the following relations 
\begin{linenomath}
\begin{subequations}
\begin{align}
    \zeta'_2(\bx,t) = \half A'_{22}\efact + \half A'_{20} + \cc,
    \\
    \Phi'_2(\bx,z,t) = \half B'_{22}\efact + \half B'_{20} + \cc,
\end{align}
\end{subequations}
\end{linenomath}
the evolution equations for the second-order elevation $\zeta_2'$ and potential $\Phi_2'(\bx,0,t)$ can be derived, which conform with \specialeqref{eq:bc_sf_prime}{a,b}. 

Again, \specialeqref{eq:AB_22}{a,b} and \specialeqref{eq:AB_20}{a,b} in the framework of CEEEs suggest that the computational accuracy and efficiency for numerical solutions relies only on the spatial and temporal scales of the nonlinear forcing terms $\mn_A\mj$ and $\mn_B\mj$, in a way similar to the numerical implementation of \specialeqref{eq:ABp_22}{a,b} and \specialeqref{eq:ABp_20}{a,b} by \cite{lili21}. Especially, \cite{lili21} have  demonstrated that the (second-order accurate) envelope framework due to \specialeqref{eq:ABp_22}{a,b} and \specialeqref{eq:ABp_20}{a,b} permits a significant improvement in the numerical efficiency at no cost of the accuracy for weakly nonlinear waves, compared with the numerical simulations based on the HOS method \citep{dommermuth87,west87} and the Fourier kernels \citep{hasselmann62,zakharov68,dalzell99}. Therefore, it can be conjectured that the CEEEs should have the same numerical advantage, at least up to second-order in wave steepness due to the same numerical features from comparing \specialeqref{eq:AB_22}{a,b} and \specialeqref{eq:AB_20}{a,b} from the CEEEs with \specialeqref{eq:ABp_22}{a,b} and \specialeqref{eq:ABp_20}{a,b} by \cite{lili21}, respectively.

%---%---%----%----%----%
%       section        %
%---%---%----%----%----%
%
\subsection{Comparisons with the HOS method}
\label{sub:hoscomp}
The CEEEs are next compared with the HOS method in the perspective of the numerical implementation of the nonlinear forcing terms (\S\ref{sub:num_nonlinear}) in a limiting case and of the computational efficiency for the same level of accuracy (\S\ref{sub:comp}). It is especially noted that the relation between the HOS method and the Zakharov integral equation has been shown in \cite{onorato07} and thereby the relation between the CEEEs and the Zakharov integral equation can be established through the HOS method. 
\subsubsection{Comparisons of the nonlinear forcing terms}
\label{sub:num_nonlinear}
\begin{figure}
	\centering	\includegraphics[width=1\columnwidth]{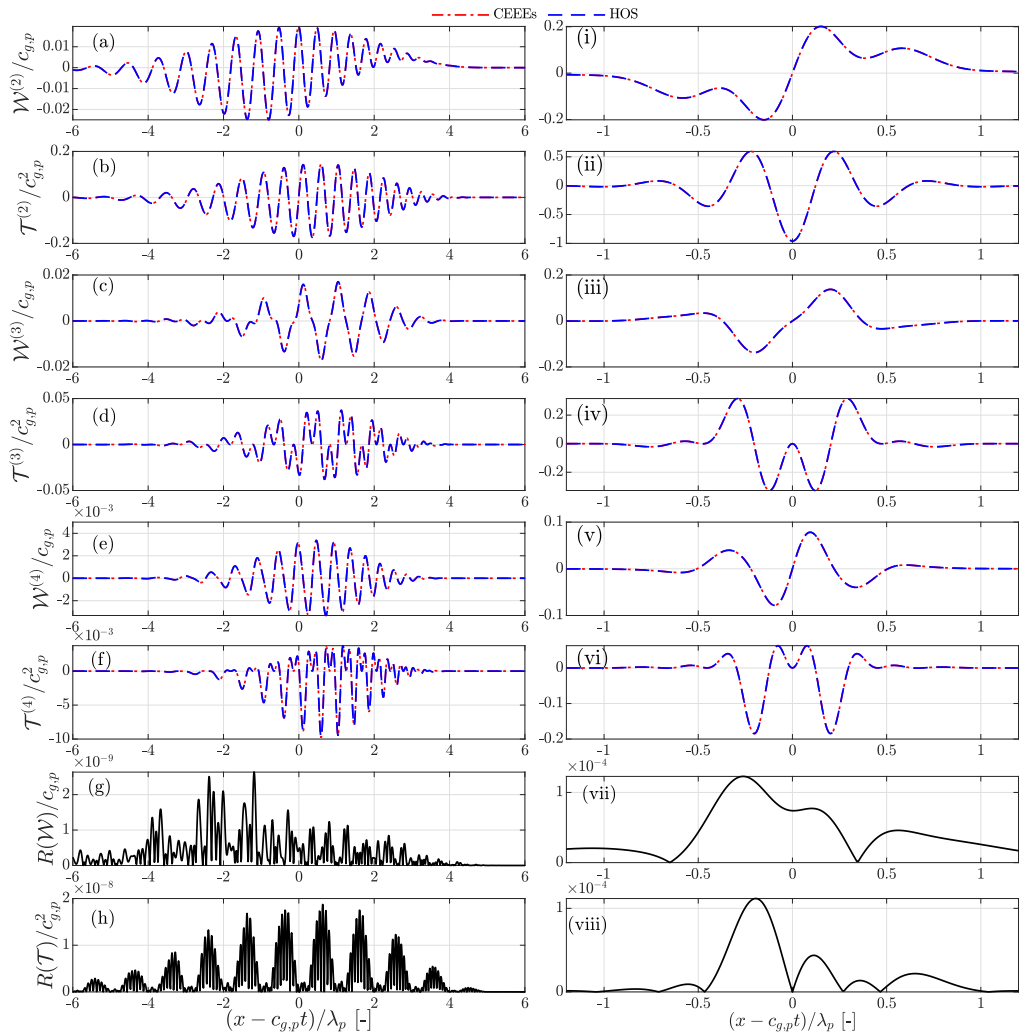}
    \caption{Comparisons of the nonlinear forcing terms (panels (a--f, i--vi)) at different orders in wave steepness, and the differences of the nonlinear forcing terms $\mw_{M}$ and $\mt_{M}$ for $M=4$ (panels (g,h, vii, viii)),  between the HOS method (blue dashed) and the CEEEs-based model (red dot-dashed) based on the equations presented in \S\ref{sub:hos_ss} and \S \ref{sub:cees_wt}, respectively. (a-h) $t=-15\times T_p$ and (i-viii) $t=0\times T_p$ for the wavepacket at the linear focus, with $T_p$, $c_{g,p}$ and $\lambda_p$  the period, group velocity, and wavelength of the spectral peak wave of a JONSWAP spectrum, respectively; $k_pA_\tf=0.8$ and $k_ph=1.5$ were used, where $k_p=2\pi/\lambda_p$, $h$ is the water depth, and $A_\tf$ is the amplitude of the focus wave at linear focus; $R(\chi)$ denotes the absolute differences of an arbitrary field $\chi$ obtained from CEEEs and HOS method, respective.
 }
	\label{fig:hos_ceees}
\end{figure}
\begin{figure}
    \centering
    \includegraphics[width=1\columnwidth]{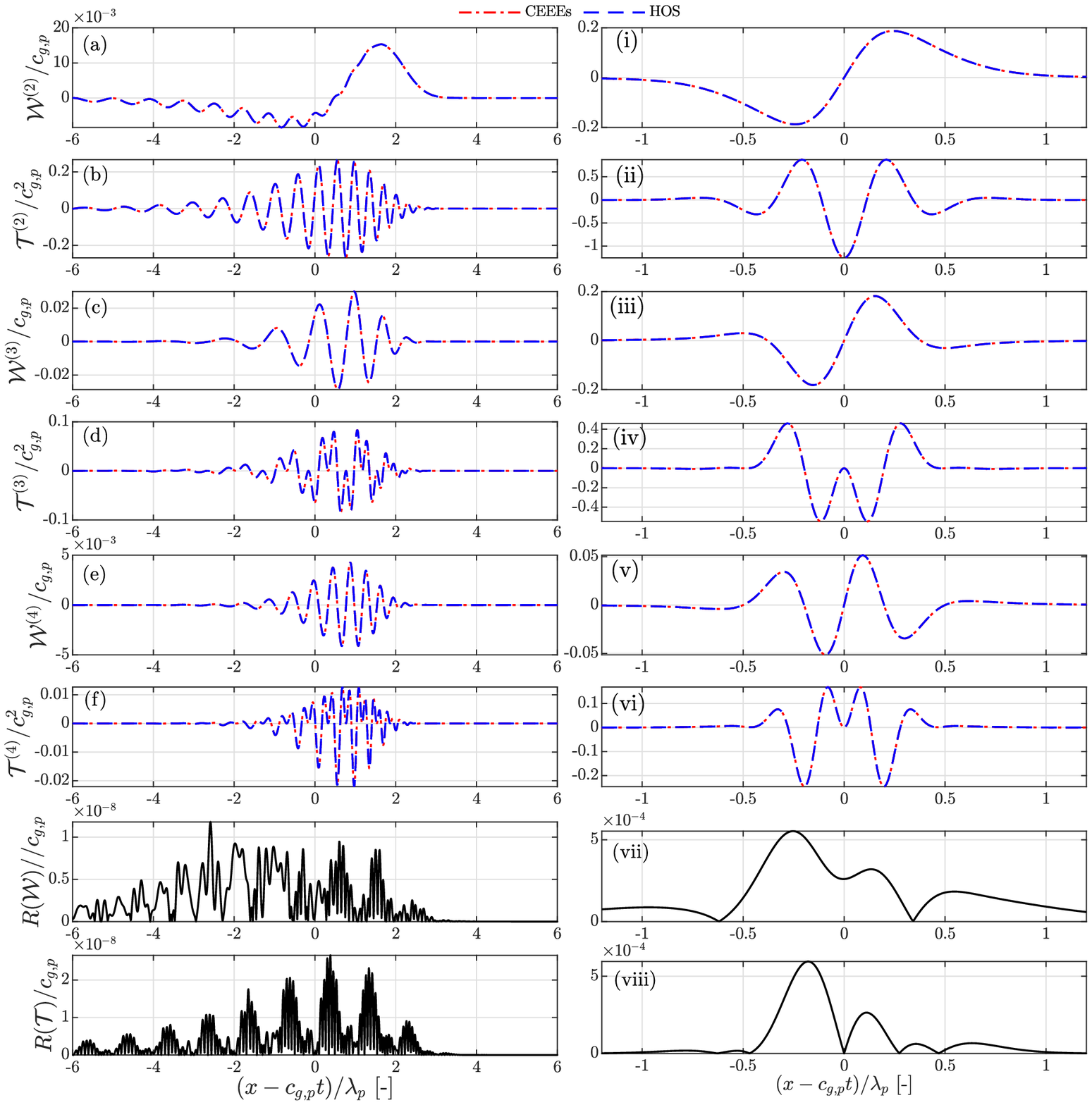}
    \caption{Caption is the same as figure \ref{fig:hos_ceees} but $k_ph=3$. }
	\label{fig:apdix}
\end{figure}

As noted above, the differences between the CEEEs and the HOS method  mainly arise from the different choices of the main unknowns that are solved for numerically. They should  in principle  return exactly the same numerical results if the  initial conditions are the same. As the detailed derivations in  \S\ref{sec:ceees} has analytically demonstrated this point, we choose to keep the numerical validation to the minimum to avoid unnecessary repetition.  In particular, we focus on that the nonlinear forcing terms given by \eqref{eq:mwm_hdef} and \eqref{eq:mtm_hdef} should be identical to \specialeqref{eq:mwt_defs}{a} and \specialeqref{eq:mwt_defs}{b} at different orders (i.e. different values of $m$) in wave steepness. The HOS method and  the CEEEs shall require  different parameters as input, the former of which uses $\zeta(\bx,t)$ and $\psi(\bx,t)$ whereas the latter $A$ and $B_s$. The relation given by \specialeqref{eq:ABs_def}{a} between $\zeta$ (or $\psi$) and $A$ (or $B_s$) should hold for all the times if both methods are used for the computation of the same case.  

A case of a right-propagating (i.e. in the positive $x$ direction) focused wave group  in two dimensions was chosen to show the comparisons between the two methods.  Especially, the parameters for the HOS method at a time instant were prescribed according to 
\begin{equation}\label{eq:zetapsi_jons}
    \zeta = ~ \mR\left[A_p\int \dfrac{|\hzt(\omega_g)|}{\sqrt{m_0}} E(x,t)\rmd \omega_g\right] 
    ~\text{and}~
    \psi =  ~\mR\left[ A_p\int \dfrac{-\rmi g|\hzt(\omega_g)|}{\omega_g\sqrt{m_0}} 
    E(x,t)
      \rmd\omega\right],
      \specialnumber{a,b}
\end{equation}
where the factor $E(x,t)=
\exp \left\{ \rmi [k (x-x_\tf)-\omega_g(t-t_\tf)+\theta_\tf]\right\}$, 
with $A_p$, $x_\tf$, $t_\tf$, and $\theta_\tf$ the prescribed peak (real) amplitude, position, time, and phase for the wave group at linear focus, respectively;  $k$ and $\omega_g$ denote the wavenumber and angular frequency of a train of right-propagating monochromatic wave, respectively, obeying the linear relationship $\omega_g=\omega(k,h)$ as defined in \S\ref{sub:2operators};  $m_0$ denotes the zeroth moment of a JONSWAP power energy spectrum $S(\omega_g)$ used with the enhancement peak factor $\gamma$ of $3.3$, and $|\hzt(\omega_g)|=\sqrt{2S(\omega_g)\Delta \omega_g}$ denotes the amplitude evaluated based on the JONSWAP with  the interval between two adjacent frequencies  prescribed on a numerical  grid. %
It is highlighted that the initial conditions given by \specialeqref{eq:zetapsi_jons}{a,b} for the initial time instant $t=t_0$ are not an exact solution to the fully-nonlinear potential-flow boundary value  problem but are the exact solution of the linearised problem. These conditions will not affect the comparisons of the nonlinear forcing terms $\mw_M$ and $\mt_M$ presented in this section between the HOS method and CEEEs as long as the initial conditions are consistent. A detailed procedure for the initialization of nonlinear waves for the solutions of an initial-value problem  can be found in works, e.g.,  \cite{dommermuth00,slunyaev16} among others.

Similarly, envelope $A$ and $B_s$ are expressed as, respectively,
\begin{equation}\label{eq:ABs_jons}
 A = ~A_p \int \dfrac{|\hzt(\omega_g)|}{\sqrt{m_0}}E_c(x,t)\rmd\omega_g
 ~\text{and}~
 B_s = ~A_p \int \dfrac{-\rmi g|\hzt(\omega_g)|}{\omega\sqrt{m_0}}  E_c(x,t)\rmd\omega_g,
 \specialnumber{a,b}
\end{equation}
with $E_c(x,t)=\exp \left\{ \rmi [(k-k_p) (x-x_\tf)-\omega_g(t-t_\tf)+\omega_pt+\theta_\tf]\right\}$, which suggest that
the following values were chosen for different parameters in the implementation of 
the CEEEs; $k_0=k_p$ and $\omega_0=\omega_p$ with $k_p=0.045$ m$^{-1}$ and $\omega_p=\omega(k_ph)$ the peak wavenumber and frequency of the JONSWAP spectrum, respectively,  $\alpha=\beta=1$, and $k_ph=1.395$.  We highlight that the elevation $\zeta$ (potential $\psi$) given by \specialeqref{eq:zetapsi_jons}{a} (\specialeqref{eq:zetapsi_jons}{b}) and the elevation envelope $A$ (the potential envelope $B_s$) given by \specialeqref{eq:ABs_jons}{a} (\specialeqref{eq:ABs_jons}{b}) should obey the equation \specialeqref{eq:ABs_def}{a} (\specialeqref{eq:ABs_def}{b}) as it indeed does. A comparison of the nonlinear forcing terms at second, third, and fourth order in wave steepness is shown in 
Figs. \ref{fig:hos_ceees} and \ref{fig:apdix}
for two different time instants using two different values for the dimensional depth, $k_ph$. The computations from the HOS method were obtained from inserting \specialeqref{eq:zetapsi_jons}{a,b} for the elevation and potential into \eqref{eq:mwm_hdef} and \eqref{eq:mtm_hdef} and the results based on the CEEEs were through substituting \specialeqref{eq:ABs_jons}{a,b} for the envelopes into \specialeqref{eq:mwt_defs}{a}  and \specialeqref{eq:mwt_defs}{b}. In the case implemented in 
Figs. \ref{fig:hos_ceees} and \ref{fig:apdix}, 
an unrealistically high value of $0.8$ for the dimensionless wave steepness $\eps=k_pA_p$  was used with the intention of demonstrating the possible differences between the two methods, if any. 
Figs. \ref{fig:hos_ceees} and \ref{fig:apdix} show
 evidently good agreement of the nonlinear terms in  all orders (in wave steepness)  between the HOS method and CEEEs. Especially, the panels in the lowest two bottom rows of Figs.\ref{fig:hos_ceees} and \ref{fig:apdix}  suggest that the absolute differences of the predictions between the two methods are a few orders of magnitude smaller than the individual predictions in the highest order in wave steepness (e.g., compared with the nonlinear forcing terms in the fourth order in wave steepness). The results shown in 
Figs. \ref{fig:hos_ceees} and \ref{fig:apdix}
are  in accordance with the theoretical derivations presented in \S\ref{sec:ceees}.

%
%---%---%----%----%----%
%        section       %
%---%---%----%----%----%
%
\subsubsection{Computational complexity}
\label{sub:comp}
We compare the numerical performances between the HOS method and CEEEs for which both are based on FFTs for the evaluation of the nonlinear terms in the evolution equations. It should be noted that the comparison is not to seek the optimal numerical methods for computations. Instead, it aims to examine three main aspects essential to numerical efficiency for achieving the same level of accuracy; (i) the number of FFTs (including inverse FFTs) needed for advancing one time step% because it is crucial to the total computational cost
; (ii) the criterion for the choice of the domain spacing and size and the time interval for numerical convergence and instability; (iii) the mathematical truncation error; see, e.g., \citet[\S1.3]{atkinson08}, which indicates the numerical accuracy that can be achieved. The comparisons of these aspects are shown in table \ref{tab:comp}. 

A total number of FFTs (including inverse FFTs) needed for updating one time step depends on a specific numerical algorithm for time integration. Therefore, in order to make the comparisons as fair as possible for the HOS method, three widely-known numerical algorithms are chosen to this end, including the first-, second-, and fourth-order accurate forward Euler (`FE1'),  mid-point (`MP2'), and Runge–Kutta (explicit, `RK4') method, respectively. The main difference between the use of an exponential integrator and the previous three algorithms lie in that a numerical algorithm for the former and the latter for time integration is carried out in the Fourier $\bk$ and physical plane, respectively, due to which the former leads to an accurate evaluation of the temporal evolution arising from the linear terms in equations. 
How to apply an exponential integrator for implementing the CEEEs is presented in \S\ref{sub:expInt}. Similar procedures can also be taken for the HOS method. An exponential integrator is used in table \ref{tab:comp} for both the HOS method and the CEEEs. An exponential integrator together with a first-order accurate numerical algorithm to approximate the time integral in the Fourier $\bk$ space (`ExpInt1') is sufficient to demonstrate the numerical features of the CEEEs, as shown in table \ref{tab:comp}, whereas additional second- (`ExpInt2') and third-order (`ExpInt3') accurate algorithms are listed for the HOS method. How to optimize the numerical implementation of the CEEEs is not the main focus of this paper and is open for studies in future works. 

We introduce a few parameters for the discussion about numerical performances. Let $L_s$, $f_s$, and $k_s$ be the  wavelength, frequency, and wavenumber 
of the shortest wave that can be represented numerically to a sufficiently good level of accuracy   with $k_sL_s=2\pi$. The use of FFTs and inverse FFTs requires an evenly spaced computational domain which has a characteristic length of $L$ and the spacing of $|\Delta \bx_n|$ between two adjacent discrete points.   
The following dimensionless parameters are defined 
\begin{equation}
    \varepsilon_t = f_s\Delta t_n, ~
    \varepsilon_\tf = \dfrac{f_s-\beta Mf_0}{f_s} , ~
    \text{and}~
    \varepsilon_k =\dfrac{k_s-\alpha Mk_0}{k_s},
    \specialnumber{a,b,c}
\end{equation}
where $\varepsilon_t$ denotes the dimensionless time interval for indicating the mathematical truncation error for computing the  time integration, $f_0=\omega_0/(2\pi)$ and $\Delta t_n$ the time interval. It should be noted that $f_s\gtrsim \beta Mf_0$ and $k_s\gtrsim \alpha Mk_0$ are assumed which should hold in most practical situations as the range $0\leq \alpha\lesssim 2$ and $0\leq \beta\lesssim 2$ are recommended. Thus, the inequalities $0\lesssim \varepsilon_\tf\leq 1$ and  $0\lesssim\varepsilon_k\leq 1$ hold.

Examining the algorithms which implement the HOS method in table \ref{tab:comp}, we will find that an exponential integrator together with a numerical algorithm for the time integration seems to have advantages in the numerical performances against using the other three methods. It is shown in table \ref{tab:comp} that the first-order accurate forward Euler algorithm requires the time interval to be extremely small for achieving a sufficient level of accuracy. As a result, the additional computational cost needed due to a larger number of time steps would not be made up by the computational efficiency saved from a smaller number of FFTs required per step, compared with the other three methods. For weakly nonlinear waves where the wave steepness $\eps\to 0^+$ and $\varepsilon_t\ll 1$ for convergent numerical results, and therefore the assumption of the scales $\mo(\eps)\sim \mo(\varepsilon_t)$ can be made,  the computational efficiency of an exponential integrator (i.e., `ExpInt1' and `ExpInt2') is superior to the mid-point method (`MP2') due to either a less number of FFTs or a higher accuracy when the other aspects are kept the same. For steeper waves where the wave steepness can be one order of magnitude larger compared to weakly nonlinear waves, and thus $\mo(\eps)\sim \mo(\sqrt{\varepsilon_t})$ can be assumed, the same conclusion can be drawn for a second-order accurate exponential integrator (`ExpInt2') as it leads to obviously a higher level of accuracy despite of a slightly increased cost arising from a larger number of FFTs. Compared with the Runge-Kutta algorithm (`RK4'), the third-order accurate  exponential integrator (`ExpInt3')  shows a slightly better performance due to a smaller number of FFTs required for achieving the same level of accuracy with $\mo(\eps)\sim\mo(\varepsilon_t)$.  

% -------------------------------------------------------------
\begin{table}
	\caption{Comparison of the numerical performances for convergent and stable computations between the HOS method and the CEEEs. The columns below the order (i.e. $M$) of accuracy in wave steepness ($\eps$) show the total number of FFTs (including  inverse FFTs) needed for advancing one time step. In the table, $f_s$ and $k_s$ denote the  frequency and wavenumber of the shortest wave that can be represented numerically to a sufficiently good level of accuracy, respectively; $N_s$ denotes the selected number of points used per characteristic wavelength; $\eps$, $\varepsilon_\tf = (f_s-M\beta f_0)/f_s  $ ($0\lesssim \varepsilon_\tf\leq 1$), and $\varepsilon_k=(k_s-M\alpha k_0)/k_s $ ($0\lesssim\varepsilon_k\leq 1$) denote the dimensionless wave steepness, and characteristic dimensionless bandwidth in frequency and wavenumber, respectively, with $f_s\gtrsim\beta Mf_0$, $k_s\gtrsim\alpha Mk_0$, 
	% $f_s=M\beta f_0/(1-\varepsilon_t)$,  
	$k_s=M\alpha k_0/(1-\varepsilon_k)$, and $\alpha$ and $\beta$ arbitrarily chosen {\it non-negative parameters}.Different choices of approximate methods were made for the time integration in the numerical implementation of the HOS method;  `FE1', `MP2' and `RK4' denotes the first-, second-, and fourth-order accurate forward Euler,  mid-point, and Runge–Kutta method for the time integration, respectively, and `ExpInt1', `ExpInt2', and `ExpInt3' denotes  a first-,  second-, and third-order  accurate approximate method are chosen for the time integration based on an exponential integrator, respectively. 
	% {\color{red} This table still needs to a double check!}
	}
	\centering
	\begin{tabular}{cccccccc}
		\hline
		Method  & \multicolumn{3}{c}{Order of} & Time                           & Domain
		& Domain length 
		& Error from
		\\
		(`Approximation to &  \multicolumn{3}{c}{accuracy$(M)$}   &  interval           &  spacing            
		&   ($L$)  with $N_k$
		& mathematical
		\\
		time integration')& 2              & 3              & 4 		           & ($\varepsilon_t\equiv f_s\Delta t_n$)            & ($\Delta \bx_n$)              
		&   Fourier modes
		& truncation
		\\
		\hline
		HOS (`FE1')  & 7              & 11             & 16             & $  \varepsilon_t  \ll 1$                     & $\dfrac{|\Delta\bx_n|}{L_s} \sim \dfrac{1}{N_s}$         
		& $\dfrac{L}{|\Delta x_n|}\sim N_k$
		& $\mo(\varepsilon^2_t)$
		\\
		HOS (`MP2')  & 14              & 22             & 32            & $  \varepsilon_t \ll 1$                     &   $\dfrac{|\Delta\bx_n|}{L_s} \sim \dfrac{1}{N_s}$        
		& $\dfrac{L}{|\Delta x_n|}\sim N_k$
		& $\mo(\varepsilon^3_t)$
		\\
		HOS  (`RK4') & 28             & 44             & 64            & $  \varepsilon_t \ll 1$                     & $\dfrac{|\Delta\bx_n|}{L_s} \sim \dfrac{1}{N_s}$    
		&  $\dfrac{L}{|\Delta x_n|}\sim N_k$
		& $\mo(\varepsilon^5_t)$
		\\
		HOS  (`ExpInt1') & 9              & 13             & 18             & $  \varepsilon_t \ll 1$                     & $\dfrac{|\Delta\bx_n|}{L_s} \sim \dfrac{1}{N_s}$  
		&  $\dfrac{L}{|\Delta x_n|}\sim N_k$
		& $\mo(\eps\varepsilon^2_t)$
		\\
		HOS  (`ExpInt2') & 18              & 26             & 36            & $  \varepsilon_t \ll 1$                     & $\dfrac{|\Delta\bx_n|}{L_s} \sim \dfrac{1}{N_s}$  
		&  $\dfrac{L}{|\Delta x_n|}\sim N_k$
		& $\mo(\eps\varepsilon^3_t)$
		\\
		HOS  (`ExpInt3') & 27              & 39             & 54             & $  \varepsilon_t \ll 1$                     & $\dfrac{|\Delta\bx_n|}{L_s} \sim \dfrac{1}{N_s}$  
		&  $\dfrac{L}{|\Delta x_n|}\sim N_k$
		& $\mo(\eps\varepsilon^4_t)$
		\\
		CEEEs (`ExpInt1') & 12             & 23             & 35             & $ \varepsilon_\tf \varepsilon_t \ll 1 $ & $\dfrac{|\Delta\bx_n|}{L_s} \sim \dfrac{1}{\varepsilon_k N_s}$  
		&  $\dfrac{L}{|\Delta x_n|}\sim \dfrac{N_k}{\varepsilon_k}$
		&  $\mo(\eps\varepsilon_\tf\varepsilon^2_t)$
	\end{tabular}\label{tab:comp}
\end{table}
%
% ------------------------------------------------------------%
% ------------------------------------------------------------%
%

Due to the above discussion about the HOS method examined in table \ref{tab:comp}, the comparison between the CEEEs and the HOS method will focus only on their implementation through using an exponential integrator which has been demonstrated to have favored features for the HOS method.  The numerical performances of the CEEEs shown in table \ref{tab:comp} especially depend on the choice of the values for $\alpha$ and $\beta$. Thereby, discussions are made based on the categories of the choices listed in the following, starting from the least advantageous category for the CEEEs. 
\begin{enumerate}
    \item[(a)] $\alpha=\beta=0$ and therefore, $\varepsilon_\tf=\varepsilon_k\equiv 1$. It suggests that the CEEEs are simply the HOS but with the newly introduced envelope transform for the evaluation of the nonlinear forcing terms and wave parameters. The newly proposed evaluation through the envelope transform is at the expense of an increased number of FFTs and therefore, an additional computational cost with all the other parameters  the same as the HOS method (`ExpInt1'), as clearly seen in table \ref{tab:comp}. As a result, this category of the value for $\alpha$ and $\beta$ should be dropped out if the CEEEs will be implemented for numerical computations as it does not introduce additional advantages compared with the HOS method. 
    \item[(b)] $\alpha=0 $ and $0< \beta$ and thus,  $\varepsilon_k=1$ and $0\lesssim \varepsilon_\tf< 1$, respectively. This suggests that a (small) dimensionless bandwidth parameter (i.e. $\varepsilon_\tf$) in wave frequency has been introduced. The characteristic wave frequency, $\omega_0$, can be chosen as  the angular frequency of either a carrier wave or the peak wave of a wave spectrum. A specific value for $\beta$   can be selected which permits $\varepsilon_\tf\sim \varepsilon_t$ in practice. For instance, $\beta=1$ has been used in \S\ref{sub:nls_stokes} which has led to $\varepsilon_\tf\sim\eps^2$. Table \ref{tab:comp} indicates that the first-order accurate exponential integrator (`ExpInt1') for the CEEEs can reach the same accuracy as a second-order accurate exponential integrator (`ExpInt2') for the HOS method but with a slightly smaller number of FFTs required. Moreover, the CEEEs can allow for a much larger time interval, e.g., $\varepsilon_t\sim \mo(1)$, for numerical convergence as long as the inequality $\varepsilon_\tf\varepsilon_t\ll 1 $ holds. Physically, a larger time interval (e.g., $1/f_s\ll \Delta t_n$) achieved by the CEEEs  attributes to that it depends on the rate of change of a wave spectrum which has a much slower temporal scale relative to that (i.e., $1/f_s$) of the phase of the fastest wave. This feature is especially similar to a NLS equation-based model by using a split-step method as explained in \cite{lo85}. 
    \item[(c)] $\alpha>0 $ and $\beta>0$ and thereby, $0\lesssim \varepsilon_k< 1$ and $0\lesssim \varepsilon_\tf< 1$, respectively. Compared with category (b), additional advantageous features are introduced due to a small value permitted for $\varepsilon_k$ which denotes the bandwidth in wavenumber. In contrast to $\varepsilon_k=1$, the parameter $\varepsilon_k$ of a small value for the CEEEs  suggests that a much larger domain with a length of $L\sim N_k |\Delta\bx_n|/\varepsilon_k$ can be achieved at no expenses of  computational efficiency and accuracy if the same number, $N_k$,  of Fourier modes are used. This is regardless of the choice of the time dependent parameters (e.g., $\varepsilon_\tf$ and $\varepsilon_t$). Compared with a second-order accurate exponential integrator (`ExpInt2') for the HOS method, a first-order accurate exponential integrator (`ExpInt1') implementing the CEEEs has the following three features. Firstly,  it has a slightly smaller computational cost examining the number of FFTs needed. It secondly permits a much larger temporal scale for computational instability and convergence. Thirdly, with the same Fourier modes chosen for computations, a much larger domain can be allowed. It should be highlighted that the introduction of  bandwidth parameters, $\varepsilon_k$ and $\varepsilon_\tf$, is similar to a NLS equation-based model where evolution of a slowly varying envelope is described and where the linear terms of the equation can be accurately solved with a split-step together with a pseudo-spectral method \citep{li21nls}. Both the CEEEs and a NLS equation-based model permit a relatively coarse spatial domain and larger time interval for numerics while allow for the resolving of wave phase on a prescribed computational domain. 
\end{enumerate}

Recall that the primary objective of this paper is to propose a new framework which can combine the advantages of both the HOS method and a NLS equation-based model, in the sense that the new framework can reach the same accuracy as the HOS method and, similar to a NLS equation-based method, it permits for both a larger spatial domain and slower temporal scale but at no expenses of  computational efficiency. The choice of the values for $\alpha$ and $\beta$ indicated by category (c) has indeed demonstrated that the newly derived CEEEs can indeed reach this objective. 

%-------------%---%----%----%----%--------------%
%                    section                    %
%-------------%---%----%----%----%--------------%
%
\section{Conclusions}
\label{sec:conclud}
This paper deals with the description of surface gravity waves on a finite water depth in the framework of potential flow theory. The main objective of this paper is to propose a framework that combines the merits of both the High Order Spectral (HOS) method \citep{dommermuth87,west87} and a nonlinear Schr\"{o}dinger (NLS) equation-based model \citep{zakharov68,davey74,dysthe179}. In particular, it can be as accurate as the HOS method  at no additional cost of numerical efficiency %\ed{and it can be derived up to arbitrary order in wave steepness} 
on the one hand. Similar to a NLS equation-based model, it shall be capable of describing the slow temporal evolution of a wave spectrum but at no expense of accuracy, permitting a coarse and large computational domain compared with the characteristic length of wave phase, on the other hand. To this end, a novel theoretical framework has been presented in the Hamiltonian theory based on a perturbation expansion, which leads to the coupled envelope evolution equations (CEEEs) given by \specialeqref{eq:ceees}{a,b}. The CEEEs can be derived  up to arbitrary order in wave steepness and have a much slower temporal and larger spatial scale compared with the characteristic time and wavelength of wave phase, respectively. In the new theoretical framework, the envelope of both surface elevation and the potential on the free water surface are introduced, which are shown to be a pair of canonical variables for the first time. The two envelopes are used for expressing wave fields and thereby are the main unknowns solved for numerically from the CEEEs. 

A few main features of the CEEEs are explored in this paper.  Firstly, based on the CEEEs, the energy balance equation for the evolution of wave actions is derived. Secondly, due to that the CEEEs are composed of both linear and nonlinear terms in wave steepness, it is proposed to solve the CEEEs by using an exponential integrator which leads to the analytical description of linear wave fields. Furthermore, the nonlinear terms are expressed in a form of the separation of different wave harmonics, due to which they can be especially split into two categories: one which can only force bound waves that do not obey the dispersion relation and the other which is capable of the forcing of free waves if particular conditions are met. Analytical derivations are  presented showing how the forcing of free waves can be led arising from quartet and quintet resonant interactions of linear waves. Much more physical implications remain to be explored in future works.

The newly derived framework has been compared with three different  methods. Analytical relations between the CEEEs and two traditional perturbation methods are established, including the theory for the evolution of a train of Stokes waves up to second order by \cite{fenton85} and the second-order semi-analytical framework for three-dimensional surface waves with arbitrary bandwidth and large directional spreading by \cite{lili21}. One would find that the relations established can be extended for more general cases. For example, proceeding to an order higher the CEEEs would be shown to recover a third-order accurate NLS equation in the limiting case of narrow-band waves \citep{trulsen00} or of the removal of secular terms at the third order in wave steepness \citep{li21nls}.  Through the case of the evolution of a nonlinear focus wave group examined numerically,  the nonlinear terms from the second to fourth order based on the CEEEs are shown to  be identical to these based on the HOS method. Compared with the HOS method for the same level of accuracy, the CEEEs do not require a larger number of fast Fourier transforms for computations but allow for a much larger time interval and computational domain with both a larger size and spacing. 

The new framework has potential of the applicability in more general cases which account for the interaction between surface waves and their ambient environments with different scales, e.g., sub-mesoscale currents and small-scale turbulence. Despite that the surface tension and a slowly varying water depth are neglected in the framework, an extension to additionally consider these features would be straightforward.  The new framework is in principle a spectral method and thereby it would suffer from the drawbacks due to the use of a fast Fourier transform in a way similar to the HOS method. 

%

%---%---%----%----%----%
%       section        %
%---%---%----%----%----%
%
\section*{Acknowledgement}
The author acknowledges the financial support from the Research Council of Norway through the Fripro Mobility project 287389 and POS-ERC project 342480. The author is grateful for the valuable suggestions and comments from the anonymous referees,  which have improved the quality of the paper. 

%---%---%----%----%----%
%              section             %
%---%---%----%----%----%
%
\section*{Declaration of interests}
The author reports no conflict of interest.

%---%---%----%----%----%
%              section             %
%---%---%----%----%----%
%
\appendix
%
% \section{At fifth order}
% \label{app:5th}

% \begin{align}
% B_0^{(55)} =  ~& -  \dfrac{1}{16} A^4\p_{zzzz}B -
% \dfrac{1}{16} A^3\p_{zzz}B\ssd -
% \dfrac{1}{12} A\p_{zz}B\trdt ,
% %%
% \end{align}

% %%%%
% \subsection{The vertical velocity and velocity potential on the free water surface}
% \begin{align}
% %
% \Phi^{(5)}(\bx,z,t)  = ~ & \half \Phi^{(51)}(\bx,z,t)  +\half  \Phi^{(53)}(\bx,z,t)  +\half  \Phi^{(55)}(\bx,z,t)  +\cc, 
% \end{align}
% %
% Similarly, the vertical velocity at the free water surface, $W(\bx,t)$, is expressed in the same form as $\Phi$ (i.e. \specialeqref{eq:Phiij}{a--e} ) with
% %
% \begin{align}
% W^{(i)}(\bx,t) = \sum_{j=0}^{j=i} \brW^{(ij)}(\bx,t)\rme^{\rmi j (\bk_0\cdot\bx - \omega_0t)}, 
% \end{align}
% %
% where $\brW^{(ij)}(\bx,t) = 0$ for $ij=10, ~21,~32,~41,~43,~50,~52$, and $ij=54$, 
% %
% \begin{subequations}
% 	\begin{align}
% 	\brW^{(55)}_0 =  ~& - \left( \dfrac{1}{16} A^4\p_{z}B +
% 	\dfrac{1}{16} A^3\p_{zz}B\ssd +
% 	\dfrac{1}{12} A\p_{zzz}B\trdt 
% 	\right),\\
% 	%
% 	\brW_0^{(ij)} =~ & \kint\int\limits_{-k_0}^\infty \hat{B}^{(ij)}_0(\bk,t)\efackx\rmd k_x\rmd k_y~
% 	\text{for}
% 	~ij=40,~42, ~51, ~\text{and}~53,
% 	\end{align}
% \end{subequations}

% \subsection{The nonlinear forcing terms in the CEEEs}

\section{The CEEEs to arbitrary order in wave steepness}
\label{app:Morder}
\subsection{Velocity potential and vertical velocity}
\label{app:vel}
As noted and similar to the HOS method, the CEEEs can be derived up to arbitrary order in wave steepness. This in principle relies on separating the wave harmonics in different orders in wave steepness. Recall an approximation to the wave fields given by 
\begin{equation}
    \Phi(\bx,z,t)\equiv \sum_{m=1}^M \Phi^{(m)}
    (\bx,z,t)
    ~\text{and}~
    w(\bx,z,t) =~ \sum_{m=1}^M w^{(m)}(\bx,z,t),     
    \specialnumber{a,b}
\end{equation}
where the approximate forms in different orders in wave steepness admit 
\begin{linenomath}
\begin{subequations}\label{eq:hos_ceees}
\begin{align}
    \Phi^{(m)}(\bx,z,t)  
    \equiv ~& 
    \sum_{j=0}^m \left(\half B\mj\Xi^j + \cc \right),
    \\
    w^{(m)}(\bx,z,t) \equiv \p_z\Phi^{(m)} 
    = ~& 
    \sum_{j=0}^m \left(\half \p_z B\mj \Xi^j + \cc \right),
\end{align}
\end{subequations}
where $\Xi(\bx,t;\alpha,\beta) = \exp\big[{\rmi (\alpha\bk_0\cdot\bx-\beta\omega_0t)}\big]$,  an arbitrary wave field with superscript `(10)' vanishes by definition, e.g., $B^{(10)}=0$. Due to the Laplace equation for the velocity potential in different orders in wave steepness and the seabed boundary condition, we obtain
\begin{align}
    B\mj(\bx,z,t) = \kint
    \hat{B}_0\mj(\bk,t)
    \dfrac{\cosh \big[|\bk+j\alpha\bk_0|(z+h)\big]}{\cosh \big(|\bk+j\alpha\bk_0|h\big)} 
    \efackx\rmd \bk,
\end{align}
where $\hat{B}_0\mj$ denotes the envelope of the $m$-th order velocity potential of the $j-$th harmonic evaluated at the still water surface, ${B}_0\mj\equiv B\mj(\bx,0,t)$,   transformed to the Fourier $\bk$ space, whose explicit expression remains to be derived in the following. Recall the $m-$th order velocity potential at the still water admits three equating forms; explicitly
\begin{subequations}
\begin{align}
    \Phi^{(m)}_0=  ~&  
    -\sum_{n=1}^{m-1} 
    \dfrac{1}{n!} \zeta^{n}\p^n_z\Phi^{(m-n)},
    \label{eqapp:phim0}
    \\    
   \Phi^{(m)}_0 \equiv ~&
    \sum_{j=0}^m \left(\half \Bar{\Phi}_0\mj\Xi^j + \cc \right)
    \\ \equiv ~&
    \sum_{j=0}^m \left(\half B_0\mj\Xi^j + \cc \right),
\end{align}
\end{subequations}
which include both the expressions used in the HOS method and  CEEEs. The detailed procedures for the latter are especially explained here. This relies on the explicit expression for $\bar{\Phi}_0\mj$.
It is straightforward to find that the following identities hold
\begin{subequations}\label{eq:zetan_phin}
\begin{align}
    \zeta^n \equiv ~& \left(
    \half A \Xi^j + \cc
    \right)^n   = 
       \sum_{2q\geq n}^n
    \left( \dfrac{n! }{q!(n-q)!}\dfrac{1}{2^n}
	A^q (A^*)^{n-q}\Xi^{(2q-n)} + \cc \right),
 \\
 \p_z^n\Phi^{(p)}   
     = ~& 
    \sum_{j=0}^{j=p} \left(\half \p_z^n B^{(pj)}\Xi^j + \cc \right),
\end{align}
\end{subequations}
where $q\in\{0, 1,2,...\}$  and $p\equiv m-n\in\{0, 1,2,...\}$. Inserting \specialeqref{eq:zetan_phin}{a,b} for $\zeta^n$ and $\p_z^n\Phi^{(p)} $, respectively, into \eqref{eqapp:phim0} and collecting the terms with the same power $j$ of $\Xi$ leads to the expression for $\bar{\Phi}\mj_0(\bx,t)$: 
\begin{align}\label{eqs:BW_mj}
    %  B\mj_0 
   \Bar{\Phi}\mj_0   = ~&      
      - \sum_{n=1}^{m-1}\sum_{2q\geq n}^{q= n}
      \sum_{r=0}^{m-n}
      \left\{\dfrac{1}{q!(n-q)!}\dfrac{1}{ 2^{n}} 
      \left[
 A^q (A^*)^{n-q}\right.\right. 
 \times
 \notag\\ 
 & \left(\delta_{2q-n-r,j}  \big(\p^n_z B^{(pr)}\big)^*
 +\ed{\text{sgn}(j)} \delta_{2q-n+r,j}\p^n_z B^{(pr)}
 \right) + 
 \notag
 \\
 & ~~~~\left. \left.
 \ed{\text{sgn}(j(2q-n)) }
 \delta_{r-2q+n,j} (A^*)^q A^{n-q}\p^n_z B^{(pr)}
 \right]%\Xi^j
  %  +\cc
   \right\},
\end{align}
\end{linenomath}
where $\delta_{i,  j}$ is the Kronecker delta function which returns unity for $i=j$, or zero otherwise\ed{; sgn$(z)$ denotes the signum function}. Based on \eqref{eqs:BW_mj}, one wound find that the velocity potential given by \eqref{eqs:BW_mj} vanishes for $mj=\{21, 41, 43, ...\}$ arising from the definition of vanishing envelopes for $mj=10$. The envelopes $B_0\mj$, are then obtained based on  $\bar{\Phi}_0\mj$ through the identity given by \eqref{eqs:hBmj} for $m\neq j$ and by \eqref{eqs:Bmm} for $m=j$. 

Similarly, the vertical velocity at the still-water surface is given by
\begin{linenomath}
\begin{subequations}
\begin{align}
    W(\bx,t) =~& \sum_{m=1}^M W^{(m)}(\bx,t)
    ~\text{with}~\\
    W^{(m)}(\bx,t) = ~& 
    \sum_{n=0}^{m-1} \dfrac{1}{n!} \zeta^{n}\p^{n+1}_z\Phi^{(m-n)}, \label{eqapp:wcm}
    \\
   \equiv~& 
    \sum_{j=0}^m \left(\half \brwc\mj \Xi^j+ \cc \right),
\end{align}
\end{subequations}
where $\brwc\mj $ is the envelope of the $m$-th order vertical velocity at the free water surface of the $j$-th harmonic, obtained from inserting the envelopes of the velocity potential at the still water surface into \eqref{eqapp:wcm} and collecting the same power $j$ of $\Xi$; explicitly
\begin{align}
\brwc\mj = & \sum_{n=0}^{m-1}\sum_{2q\geq n}^{q= n}
      \sum_{r=0}^{m-n}
      \dfrac{1}{q!(n-q)!}\dfrac{1}{ 2^{n} } 
      \left[
 A^q (A^*)^{n-q}\right.
 \times
  \\ 
 &\left(\delta_{2q-n-r,j} 
 \right.
 \left.
 \big(\p^{n+1}_z B^{(pr)}\big)^*
 + \ed{\text{sgn}(j)}\delta_{2q-n+r,j}\p^{n+1}_z B^{(pr)}
 \right) + 
 \notag
 \\
& \left.
 \ed{\text{sgn}(j(2q-n))}  \delta_{r-2q+n,j} (A^*)^q A^{n-q}\p^{n+1}_z B^{(pr)}
 \right].\notag
\end{align}
\end{linenomath}

\subsection{Nonlinear forcing terms}
\label{sub:appforce}
Recall that the definition of the nonlinear forcing terms in the HOS method is given by
\begin{equation}
    \mw_M = \sum_{m=1}^M \mw^{(m)}
    ~\text{and}~
    \mt_M = \sum_{m=1}^M \mt^{(m)},
    ~\specialnumber{a,b}
\end{equation}
where
\begin{subequations}
\begin{align}
    \mw^{(m)} \equiv ~& \mathcal{H}(m-1.5) W^{(m)} - \delta_{m,2} \nabla\psi\cdot\nabla\zeta 
    +\mathcal{H}(m-2.5) W^{(m-2)}(\nabla\zeta)^2,
    \\
    \mt^{(m)} \equiv ~& 
    - \half\delta_{m,2} (\nabla\psi)^2 
    +\half 
 \mathcal{H}(m-1.5)\sum_{n=1}^{m-1}W^{(n)} W^{(m-n)} +
   \\
    & ~~ \half \mathcal{H}(m-3.5) (\nabla\zeta)^2\sum_{n=1}^{m-3}W^{(n)} W^{(m-n)}\notag,
\end{align}
\end{subequations}
where $\mathcal{H}$ denotes the Heaviside step function. % 
In contrast, the CEEEs propose to use the nonlinear forcing terms in different orders in wave steepness given by
\begin{equation}
    \mw^{(m)} = \sum_{j=0}^{j=m} 
    \left(\half\bmw\mj\Xi^j + \cc\right) 
    ~\text{and}~
    \mt^{(m)} = \sum_{j=0}^{j=m} 
    \left(\half\bmt\mj\Xi^j + \cc\right). 
    \specialnumber{a,b}
\end{equation}
Equating the two different expressions of the nonlinear forcing terms in the $m$-th order in wave steepness leads to 
\begin{subequations}
\begin{align} \label{eq:bmwmj_M}
    \bmw\mj = ~&
    \mathcal{H}(m-1.5) \brwc\mj - 
    \half\delta_{m,2}\delta_{j,2} (\nabla+\rmi\alpha\bk_0)B
    \cdot(\nabla+\rmi\alpha\bk_0)A 
    - 
    \\
    & 
    ~~~~~\half \delta_{m,2}\delta_{j,0}
    (\nabla+\rmi\alpha\bk_0)^*B^*
    \cdot(\nabla+\rmi\alpha\bk_0)A +
    \notag\\
    & ~~~~\half \mathcal{H}(m-2.5)\mathcal{H}(r-1.5-j)
|(\nabla+\rmi\alpha\bk_0)A|^2\brwc^{(rj)}+ 
    \notag\\
    &
    ~~~~\quard
    \mathcal{H}(m-2.5) \mathcal{H}(j-2.5)
    \left(
    [(\nabla+\rmi\alpha\bk_0)A]^2
    \right)\brwc^{(r\gamma)},
    \notag
\end{align}
where $r=m-2$ and $\gamma=j-2$ for $j\geq 3$; and 
\begin{align}\label{eq:bmtmj_M}
    \bmt\mj =~& 
    \quard\delta_{m,2}\delta_{j,2} \big[(\nabla+\rmi\alpha\bk_0)B\big]^2
    + 
    \quard \delta_{m,2}\delta_{j,0}
    \big|(\nabla+\rmi\alpha\bk_0)B\big|^2 + 
    \quard \mathcal{H}(m-1.5) \times
    \notag
    \\
    %%%%%%
    \sum_{n=1}^{m-1}\sum_{p=0}^n &
    \sum^{m-n}_{\gamma=0}
    \left[
    \delta_{p+\gamma,j}\brwc^{(np)}\brwc^{(q\gamma)}
    + \delta_{p-\gamma,j}\brwc^{(np)}\brwc^{(q\gamma)*}
    + \delta_{p-\gamma,-j}
    \brwc^{(np)*}\brwc^{(q\gamma)}  
    \right]+ 
    \notag\\
    %%%%%%%%%
    & ~~~~\dfrac{1}{16}
    \mathcal{H}(m-3.5) \mathcal{H}(j-2.5)
    \left[(\nabla+\rmi\alpha\bk_0)A\right]^2
    \times
    \\
    \sum_{n=1}^{m-3}
    \sum_{p=0}^n &
    \sum^{m-n-2}_{\gamma=0}
    \left[
    \delta_{p+\gamma,j-2}\brwc^{(np)}\brwc^{(\nu\gamma)}
    + \delta_{p-\gamma,j-2}\brwc^{(np)}\brwc^{(\nu\gamma)*}
    +
    \delta_{p-\gamma,-(j-2)}
    \brwc^{(np)*}\brwc^{(\nu\gamma)}  
    \right] + 
    \notag\\
    %%%%%%%%
   & \dfrac{1}{16}
    \mathcal{H}(m-3.5)
    |(\nabla+\rmi\alpha\bk_0)A|^2
    \times
    \notag\\
    \sum_{n=1}^{m-3}\sum_{p=0}^n &
    \sum^{m-n}_{\gamma=0}
    \left[ \delta_{p+\gamma,j}\brwc^{(np)}\brwc^{(\nu\gamma)}
    + \delta_{p-\gamma,j}\brwc^{(np)}\brwc^{(\nu\gamma)*}
    +\delta_{p-\gamma,-j}
    \brwc^{(np)*}\brwc^{(\nu\gamma)}   
    \right], \notag
\end{align}
\end{subequations}
where $p=m-n$ is noted and $\nu=m-n-2$.  Inserting \eqref{eq:bmwmj_M} and \eqref{eq:bmtmj_M} into \specialeqref{eq:mwt_defs}{a,b}, and thereby the nonlinear forcing terms on the right-hand side of the CEEEs presented in \S\ref{sub:ceees}, the CEEEs up to arbitrary order in wave steepness are therefore obtained. It shall be noted that, despite the cumbersome expressions presented in \S~\ref{app:vel}, many involved terms would vanish and there are only a very few terms that contribute to the CEEEs, especially these correct to the lowest orders in wave steepness as shown in \S\ref{sec:ceees}.

\bibliographystyle{jfm}
\bibliography{jfmref}

\end{document}